\documentstyle[11pt,aaspp4,flushrt,epsfig]{article}
\begin{document}
%
% abbreviations of useful commands
\renewcommand{\vec}[1]{\mbox{\boldmath $\displaystyle #1$}} %boldface
							    %vector style 
\newcommand{\grad}{\vec{\nabla}} %gradient
\newcommand{\divr}{{\rm div}\,}	%divergence
\newcommand{\curl}{{\rm curl}\,} %curl
\newcommand{\lap}{\nabla^2}	%laplacian
\newcommand{\vdot}{\mbox{\boldmath $\cdot$}} %vector inner product
\newcommand{\vcross}{\mbox{\boldmath $\times$}}	%vector cross product
\newcommand{\ee}[1]{\times10^{#1}} %scientific notation
\newcommand{\nuc}[1]{$\rm #1$}	%nuclear reaction notation
\newcommand{\ddt}[1]{\frac{\partial #1}{\partial t}} %partial time derivative
\newcommand{\DDt}[1]{\frac{d #1}{dt}} %total time derivative
\newcommand{\kB}{k_B}		%Boltzman's constant
\newcommand{\K}{\,{\rm K}}
\newcommand{\gram}{\,{\rm g}}
\newcommand{\secd}{\,{\rm s}}
\newcommand{\yr}{\,{\rm yr}}
\newcommand{\Myr}{\,{\rm Myr}}
\newcommand{\Gyr}{\,{\rm Gyr}}
\newcommand{\cm}{\,{\rm cm}}
\newcommand{\km}{\,{\rm km}}
\newcommand{\cms}{\cm\secd^{-1}} %velocity
\newcommand{\gpcc}{\gram\cm^{-3}} %mass density
\newcommand{\erg}{\,{\rm erg}}
\newcommand{\MeV}{\,{\rm MeV}}
\newcommand{\keV}{\,{\rm keV}}
\newcommand{\mueff}{\mu_{\rm eff}} %effective mean molecular weight
\newcommand{\fscr}{f_{\rm scr}}	%screening factor
\newcommand{\msun}{M_{\odot}}	%solar mass
\newcommand{\rsun}{R_{\odot}}	%solar radius
\newcommand{\Lsun}{L_{\odot}}	%solar luminosity
\newcommand{\Teff}{T_{\!\rm eff}} %effective temperature
\newcommand{\tdepl}{t_{\rm depl}} %depletion time
\newcommand{\NA}{N_{\!A}}	%Avogadro's constant
\newcommand{\mcoul}{0.072\,\msun} %Coulomb mass
\newcommand{\mrad}{0.5\,\msun}	%radiative transition mass
\newcommand{\midt}[1]{\raisebox{1.5ex}[0pt]{#1}} %table alignment thingie

\def\figcaption[#1]#2{\begin{figure}%
	\begin{center} %
	\epsfig{file=#1,width=5in}%
	\end{center} %
	\caption{#2}%
	\end{figure}}

\title{Light Element Depletion in Contracting Brown Dwarfs and
Pre--Main-Sequence Stars}

\author{Greg Ushomirsky, Christopher D. Matzner, Edward F. Brown, 
Lars Bildsten, Vadim G. Hilliard and 
Peter C. Schroeder\altaffilmark{1}}
\affil{Department of Physics and Department of Astronomy\\ 601
Campbell Hall \\ University of California, Berkeley, CA 94720}
 \authoremail{gregus@fire.berkeley.edu, matzner@physics.berkeley.edu, 
ebrown@astron.berkeley.edu, bildsten@fire.berkeley.edu,
opaopa@physics.berkeley.edu, peters@ssl.berkeley.edu}

\altaffiltext{1}{Currently at Space Sciences Laboratory, University of
California, Berkeley, CA 94720}

\begin{center}
{\bf Accepted to ApJ}
\end{center}

\begin{abstract}
We present an analytic calculation of the thermonuclear depletion of
the light elements lithium, beryllium, and boron in fully convective,
low-mass stars.  Under the presumption that the pre--main-sequence
star is always fully mixed during contraction, we find that the
burning of these rare light elements can be computed analytically,
even when the star is degenerate.  Using the effective temperature as
a free parameter, we constrain the properties of low-mass stars from
observational data, independently of the uncertainties associated with
modeling their atmospheres and convection. Our results are in
excellent agreement with the detailed calculations of D'Antona \&
Mazzitelli (1994) and Chabrier, Baraffe, \& Plez (1996).  Our analytic
solution explains the dependence of the age at a given level of
elemental depletion on the stellar effective temperature, nuclear
cross sections, and chemical composition.  These results are also
useful as benchmarks to those constructing full stellar models.  Most
importantly, our results allow observers to translate lithium
non-detections in young cluster members into a model-independent
minimum age for that cluster.  Using this procedure, we have found
lower limits to the ages of the Pleiades (100 Myr) and Alpha Persei
(60 Myr) clusters.  Dating an open cluster using low-mass stars is
also independent of techniques that fit upper main-sequence evolution.
Comparison of these methods provides crucial information on the amount
of convective overshooting (or rotationally induced mixing) that
occurs during core hydrogen burning in the 5--10~$\msun$ stars
typically at the main-sequence turnoff for these clusters.  We also
discuss beryllium depletion in pre--main-sequence stars.  Recent
experimental work on the low energy resonance in the
\nuc{^{10}B(p,\alpha)^7Be} reaction has greatly enhanced estimates of
the destruction rate of \nuc{^{10}B}, making it possible for stars
with $M\gtrsim 0.1\msun$ to deplete both \nuc{^{10}B} and \nuc{^{11}B}
before reaching the main sequence.  Moreover, there is an interesting
range of masses, $0.085\msun\lesssim M\lesssim 0.13\msun$, where boron
depletion occurs on the main sequence in less than a Hubble time,
providing a potential ``clock'' for dating low-mass stars.
\end{abstract}


%\newpage

\section{ Introduction }

Due to low Coulomb barriers and purely strong-interaction reactions,
most stars are able to destroy deuterium, lithium, beryllium, and
boron via proton capture prior to reaching the main sequence.  For
low-mass stars ($M\lesssim\mrad$; the exact mass cutoff depends on the
element, see \S~\ref{coreconv}), light element depletion occurs while
the star is fully convective (i.e., before the stellar core becomes
radiative) and still contracting towards the main sequence.  In fully
convective stars, light elements are depleted throughout the star, and
the photospheric abundance is a reliable indicator of the interior
abundance.  In the case of lithium, the depletion at the photosphere
in these stars is readily observable.  Deuterium is depleted on a
contraction timescale when the central temperature is $T_c\approx
(8$--$10)\times 10^5\K$, in which case the energy released from
consuming all the deuterium (${\rm D/H}\approx 1.6\times 10^{-5}$ in
the local ISM, \cite{pis97}) is comparable to the thermal energy
content in the star. This slows contraction and nearly leads to a
``deuterium main sequence'' (Grossman \& Graboske 1971, 1973; D'Antona
\& Mazzitelli 1985) at an age of order 1 Myr.  For some masses, the
deuterium burning occurs while the star is still accreting
(\cite{sta88a}).  Once accretion and deuterium burning are complete
(after 1--10 Myr, depending on the mass, Stahler 1988a) the star
undergoes traditional Hayashi contraction. It is during this phase
that lithium, beryllium, and boron are destroyed in most low-mass
stars.

There has been much previous work on light element depletion in
gravitationally contracting pre--main-sequence stars (Hayashi \&
Nakano 1963; Weymann \& Moore 1963; Ezer \& Cameron 1963; Bodenheimer
1965, 1966; Nelson, Rappaport, \& Chiang 1993; D'Antona \& Mazzitelli
1994 [hereafter DM94]; Chabrier, Baraffe, \& Plez 1996 [hereafter
CBP96]; Chabrier \& Baraffe 1997 [hereafter CB97]).  Motivated by
lithium observations along the main sequence in open clusters (see
Soderblom 1995 and \cite{mar98} for recent reviews), many have
calculated lithium depletion for stars of $M\gtrsim\mrad$ both during
and after the pre--main-sequence phase (Vauclair et al.\ 1978;
D'Antona \& Mazzitelli 1984; Proffitt \& Michaud 1989; Vandenberg \&
Poll 1989; Swenson, Stringfellow, \& Faulkner 1990; Deliyannis,
Demarque, \& Kawaler 1990; \cite{for94}; \cite{mar96}).  For stars
with $M\gtrsim\mrad$, most of the lithium burns after convection has
halted in the stellar core. Accurate predictions of lithium depletion
then depend on the temperature at the bottom of the retreating
convective zone. The location of the convective/radiative boundary and
the amount of mixing across it depend on opacity, treatment of
convection, and rotation; proper handling of these effects remains an
open question.  The range of observational results on lithium
depletion in young clusters also points to these effects playing an
important role for $M\gtrsim\mrad$ (\cite{mar98}).
 
In this paper, we analytically calculate the depletion of lithium,
beryllium, and boron for those stars that are always fully convective
during depletion. The interstellar abundances of these elements is low
enough that the energy released by their fusion cannot halt (nor
appreciably slow) stellar contraction.  Efficient convection
throughout the star allows us to represent it accurately as a fully
mixed $n=3/2$ polytrope, for which analytic treatment of element
depletion is tractable (Bildsten et al.\ 1997, hereafter \cite{bil97},
also see Hayashi \& Nakano 1963; Stevenson 1991; Burrows \& Liebert
1993).  Our analytic calculation gives a convenient formula for the
time and central temperature at which elemental depletion occurs, as
well as its dependence on stellar parameters, such as the mass and
composition of the star.  Most importantly, our treatment allows us to
survey the effects of the presently most uncertain aspect of stellar
modeling, the effective temperature $\Teff$ (which sensitively depends
on the treatment of the photospheric boundary conditions, opacities,
and the treatment of convection).  These uncertainties have motivated
a number of calculations (Pozio 1991; Magazz\`u, Mart\'{\i}n, \&
Rebolo 1993; Nelson, Rappaport, \& Chiang 1993; DM94; CBP96; CB97),
each employing different treatments of these uncertain factors.  We
show that for a given $\Teff$ and $M$ we can reproduce the results of
these prior works to high accuracy.  Moreover, our approach allows the
inferred effective temperature of an observed star to be used directly
in analyzing light element depletion observations, without first
placing it on a theoretical evolutionary track.

A previous application of our theoretical results to low-mass stars in
the Pleiades (\cite{bil97}) confirmed in a model-independent way the
100 Myr minimum age originally determined from lithium observations by
Basri et al.\ (1996).  As we discuss in \S \ref{alphaper}, applying
our method to the lithium non-detections in pre--main-sequence stars
in the open cluster Alpha Persei (Zapatero Osorio et al.\ 1996) yields
a minimum age of 60 Myr for that cluster, independently of the
effective temperature scale for low-mass stars. We stress that lithium
observations of contracting pre--main-sequence stars in young open
clusters provide accurate and reliable minimum ages.  This is very
important, because for the age range (10--200 Myr) when low-mass stars
($0.09 \lesssim M/\msun \lesssim 0.4$) are actively depleting lithium,
stars with masses of 4--15 $\msun$ are leaving the main sequence
(using Maeder \& Meynet's 1989 $t_{\rm MS}(M)$ relations with
overshoot of $1/4$ of the pressure scale height at the
convective/radiative boundary).  Confirming or denying that these
stars live longer on the main sequence due to enhanced mixing (from
convective overshoot) is a crucial issue.  White dwarfs have also been
found in many of these young clusters and extrapolating backwards in
time from the cooling age allows one to estimate the mass of the
progenitor star. This extrapolation depends critically on the presumed
age of the cluster (see Reid 1996 for an up-to-date discussion).

We begin in \S~\ref{Structure} by justifying the treatment of the
contracting star as an $n=3/2$, fully mixed polytrope.  We also
outline two important limitations to our approach, the formation of a
radiative core and the onset of substantial hydrogen burning.  Our
analytic calculation of hydrogen burning also yields a convenient
fitting formula for the stellar radius and surface gravity as a
function of $\Teff$ along the zero-age main sequence (summarized in
Appendix \ref{s:ZAMSfits}).  In \S~\ref{CaptureRates}, we update the
proton capture rates for the light elements and discuss the possible
effects of a sub-threshold resonance in proton captures on beryllium.
Section \ref{formalism} presents the equation describing light element
depletion, our treatment of the correction due to electron degeneracy
at the time of depletion, and our method for solving the resulting
equations.  In \S~\ref{results}, we provide detailed fitting formulae
for the depletion of different light elements and compare our results
to previous numerical work.  We emphasize that our technique makes
robust predictions, even when the instantaneous depletion time is much
less than the contraction time.  This is important, since lithium is
detectable even when substantially depleted.  We also apply our
results to lithium observations in open clusters.  We conclude in
\S~\ref{Conclusions} with a brief summary and a few speculations.

\section{The Internal Structure of Pre--Main-Sequence Stars}\label{Structure}
 
Detailed numerical evolutionary calculations of contracting
pre--main-sequence (and pre--brown-dwarf) stars have been carried out
by a number of authors (see Burrows \& Liebert 1993 for a review).  As
shown by Nelson, Rappaport, \& Joss (1986), very low-mass stars
($M\lesssim 0.03\msun$) remain completely and efficiently convective
even at ages of about 4.5 Gyr. Higher-mass stars ($M\gtrsim 0.07
\msun$) stop convecting only once they are degenerate enough (and,
likewise, cool enough) for electron conduction to carry the necessary
flux; this takes at least 5 Gyr (\cite{ste91}).  Of those that burn
hydrogen, stars of mass $M\lesssim 0.3\msun$ remain fully convective
even after they settle onto the main sequence, while stars of mass
$M\lesssim0.5\msun$ remain fully convective at least until the end of
\nuc{^7Li} burning (DM94).  This critical mass is lower for beryllium
and boron burning.  As long as we restrict our attention to masses
below this critical value (derived in \S~\ref{coreconv}), full and
efficient convection prevails during the period of light element
depletion.

The effective temperature of a fully convective star determines the
contraction rate and is found by matching the entropy in the deep
interior to that below the photosphere (see Stahler 1988b for an
excellent review of Hayashi contraction). For these low-mass stars
with $\Teff\lesssim 4000\K$, the opacities are still uncertain, and
the treatment of convection near the photosphere is still open to
debate.  The differing input physics results in different effective
temperatures for the same stellar mass.  For example, DM94 found that
for a $0.2\,\msun$ star, $\Teff$ ranges from 3350 K to 3640 K
depending on whether Alexander et al. (1989) or Kurucz (1991)
opacities were used in the stellar envelope.  On the other hand, for
the same star, CB97 get $\Teff=3250$~K using Alexander \& Fergusson
(1994) opacities, a non-gray atmosphere (Allard \& Hauschildt 1997),
and a more consistent treatment of convection near the photosphere, as
opposed to a gray atmosphere used by DM94.  In addition, the
uncertainty in the effective temperature inferred from the observed
spectrum can be as large as $\pm 125 \K$ (\cite{kir93}), and
discrepancies between effective temperature scales from different
authors can be much higher, up to $500 \K$ (Mart\'{\i}n, Rebolo,\&
Magazz\`u 1994; \cite{zap96}).

Despite the variations in the exact value of $\Teff$ that stem from
differences in the input physics, most calculations do agree that
$\Teff$ {\it remains approximately constant during the fully
convective contraction phase.}   Thus, rather than rely on a
particular choice of input physics to fix $\Teff(M,R)$, we choose to
view it as a free parameter that is constant during fully convective
contraction, and explore its effect on elemental depletion in a
model-independent way.

\subsection{ Fully Convective Stars}\label{structure}

In a fully (i.e., core to photosphere; see below for a caveat) and
efficiently convecting star the entropy per unit mass is independent
of position within the star and decreases as the star contracts.  This
is true even if the star is partially degenerate. The entropy of a gas
comprised of nondegenerate ions of mean molecular weight
$\mu_i\equiv\rho\NA/n_i$ and electrons with weight $\mu_e$ is 
\begin{equation}\label{eq:entropy}
s={\rm const}-\frac{\kB \NA}{\mu_i}\ln F_{1/2}(\eta)+
\frac{\kB\NA}{\mu_e}\left(\frac{5}{3}\frac{F_{3/2}(\eta)}{F_{1/2}(\eta)}-
\eta\right),
\end{equation}
where the constant term (which contains the entropy of mixing) depends
only on the composition, $\eta$ is the degeneracy parameter (ratio of
the electron chemical potential to $\kB T$), $F_n(\eta)$ is the
$n$th-order Fermi-Dirac function
\begin{equation}
F_n(\eta)=\int_0^\infty\frac{x^n
dx}{1+\exp(x-\eta)},
\end{equation}
and $F_{1/2}(\eta)\propto\rho/\mu_eT^{3/2}$ (see, for example,
\cite{cla83}).  The entropy is a monotonic function of the degeneracy
parameter, so for constant $\mu_e$ and $\mu_i$ equation
(\ref{eq:entropy}) is invertible, $\eta = \eta(s)$.  Therefore, in an
isentropic star the degree of electron degeneracy is constant
(Stevenson 1991; Burrows \& Liebert 1993; \cite{bil97}).  The equation
of state is
\begin{equation}\label{eos-degenerate}
P=\frac{\rho \NA}{\mueff}\kB T,
\end{equation}
where
\begin{equation}\label{eq:mueff}
\frac{1}{\mueff}\equiv\left(\frac{1}{\mu_i}+\frac{1}{\mu_e}\frac{2}{3}
\frac{F_{3/2}(\eta)}{F_{1/2}(\eta)}\right)\approx
\left(\frac{1}{\mu_i}+\frac{1}{\mu_e}
\frac{(1+0.1938F_{1/2}(\eta))^{5/3}}{1+0.12398F_{1/2}(\eta)}\right). 
\end{equation}
The approximation used in the second equality is from Larson and
Demarque (1964) and is accurate to within $0.02\%$ for $\eta<30$. In
essence, for a given pressure and density, the temperature is reduced
by a degeneracy dependent factor that is constant along an adiabat.
This means that the adiabatic exponents are
$\Gamma_1=\Gamma_2=\Gamma_3=5/3$, and the star is described by an
$n=3/2$ polytrope regardless of the degree of electron degeneracy
(Hayashi \& Nakano 1963; Stevenson 1991; Burrows \& Liebert 1993;
\cite{bil97}).  Moreover, $\mueff$ is constant throughout the star.
Accordingly, we have $T=T_c\Theta(\xi)$ and $\rho = \rho_c
\Theta^{3/2}(\xi)$, where $\xi \equiv \xi_1 r/R$ is the radial
coordinate, and $\Theta(\xi)$ is the Lane-Emden function for $n=3/2$,
which crosses zero at $\xi_1 = 3.65$.  The central density and
temperature are
\begin{eqnarray}
\label{rho_c-polytrope}
 \rho_c &=& 8.44 \left(\frac{M}{\msun}\right)
\left(\frac{\rsun}{R}\right)^{3} \gram\cm^{-3}, \\
\label{T_c-polytrope}
T_{c} &=& 7.41\times 10^6 
\left(\frac{\mueff}{0.6}\right)
\left(\frac{M}{\msun}\right)
\left(\frac{\rsun}{R}\right) \K.
\end{eqnarray}
In the non-degenerate limit $1/\mueff\rightarrow 1/\mu_i+1/\mu_e\equiv
1/\mu$, and the stellar model is completely specified.  For an
arbitrary degree of degeneracy, however, the stellar radius, $R$, and
$\mueff$ are related by
\begin{equation}\label{R-degeneracy}
\frac{R}{\rsun}=\frac{7.73\ee{-2}}{\mueff\mu_e^{2/3}F_{1/2}^{2/3}(\eta)}
\left(\frac{\msun}{M}\right)^{1/3}.
\end{equation}
We will use this relation in \S~\ref{formalism} to calculate the
depletion.  

In the above discussion we assumed that efficient convection assures
that the star is isentropic throughout, and the entire star is
well-described by an $n=3/2$ polytrope.  Electron degeneracy alone
cannot modify the polytropic structure to affect this relation.
However, near the photosphere, the entropy cannot be constant for two
reasons.  First, near the photosphere and above it, radiative
transport must carry most of the energy flux, and hence the entropy in
such regions must increase outwards.  Moreover, immediately below the
photosphere, and especially in the hydrogen ionization zone,
convection tends to be inefficient and superadiabatic, leading to a
decrease of entropy with radius; see Figure 3 of Stahler (1988b) for
an illustration of this behavior.  Therefore, the relation between the
entropy in the interior and that at the photosphere is not trivial.  A
significant fraction of the effort in numerically simulating
pre--main-sequence evolution is aimed at disentangling this
relationship in order to self-consistently calculate the effective
temperatures for these stars.

The effect of these complications on the stellar radius is extremely
small.  The radiative skin in the stars under consideration is
tremendously thin, $\Delta R/R < 10^{-4}$, $\Delta M\sim 10^{-10}
M_\odot$ (\cite{bur89}).  Similarly, the ratio of the temperature at
the ionization zone ($T_{\rm ioniz}\sim 10^4\K$) to the central
temperature ($T_c\sim 10^6\K$) is of order $10^{-2}$; the virial
theorem then implies that the ionization zone is located in the outer
$1\%$ of the stellar radius. Moreover, the ratio of the pressure at
the ionization zone to the central value is $P_{\rm
ioniz}/P_c=\left(T_{\rm ioniz}/T_c\right)^{5/2}$.  Since the envelope
is geometrically thin, the pressure at the ionization zone is $P_{\rm
ioniz}\approx GM\Delta M/4\pi R^4$, while the pressure at the center
of an $n=3/2$ polytrope is $P_c=9.7GM^2/4\pi R^4$, so the mass of
neutral gas in the star is just $\Delta M\sim 10^{-4} M$.  Thus, most
of the mass (and volume) of the star is fully ionized, as well as
fully (i.e., core to photosphere) and efficiently convective, and the
stellar structure is well-described by an $n=3/2$ polytrope until
radiative heat transport at the stellar center (\S~\ref{coreconv}) or
Coulomb corrections (\S~\ref{limitat}) become important.  The
difference between the true radius of the star and that given by
equations (\ref{rho_c-polytrope}) -- (\ref{R-degeneracy}) is extremely
small, and we are justified in neglecting it.

Except during deuterium burning, the energy output from burning light
elements is insufficient to halt the gravitational contraction
(however, see \S \ref{coreconv} for discussion of effects of hydrogen
burning). Thus, we assume that the luminosity of the star, $L$, is
balanced by energy generation due to gravitational contraction alone.
For an $n=3/2$ polytrope, a virial analysis gives
\begin{equation}\label{eq:L-virial}
L = 4\pi R^2 \sigma_{\rm SB}\Teff{}^4 = -\frac{3}{7} \frac{G
M^2}{R^2} \frac{dR}{dt};
\end{equation}
this relation holds even if the star is partially or strongly
degenerate. As discussed in \S~\ref{Structure}, we assume that $\Teff$
is constant during contraction. The time $t=0$ then corresponds to the
onset of quasistatic contraction, which we approximate with a
nonphysical initial state of infinite radius and contraction rate.
Therefore $t$ differs from the chronological age because of the
deuterium burning phase, the initial radius on the theoretical stellar
birthline (\cite{sta88a}), and changes in the effective temperature
during contraction. However, as long as the effective temperature is
correct when the depletions occur, $t$ will not differ significantly
from the chronological age.  Integrating equation (\ref{eq:L-virial}),
we obtain the stellar radius as a function of time,
\begin{equation}\label{R-contraction}
\frac{R}{\rsun}=\frac{17.1}{T_{e3}{}^{4/3}t_6{}^{1/3}}
\left(\frac{M}{\msun}\right)^{2/3},
\end{equation}
where $T_{e3} \equiv \Teff/1000 \K$ and $t_6 \equiv t/10^6 \yr$.
Since the radius of the star is completely determined by
$F_{1/2}(\eta)$, we combine the above relation with equation
(\ref{R-degeneracy}) to obtain a transcendental equation,
\begin{equation}\label{fhalf}
\mueff F_{1/2}^{2/3}(\eta)=4.53\ee{-3}
\left(\frac{\msun}{M}\right)\left(\frac{T_{e3}^4
t_6}{\mu_e^2}\right)^{1/3},
\end{equation}
for the degeneracy parameter as a function of time. 

\subsection{The Onset of Hydrogen Burning and the Formation of a Radiative
Core}
\label{coreconv}

As the star approaches the main sequence, p-p reactions generate an
increasing amount of energy.  Eventually, the relation for the
contraction rate (eq.\ [\ref{eq:L-virial}]), which assumes that the
energy generation is only due to gravitational contraction, ceases to
hold.  This condition defines the lower mass limit of our
calculations' validity for beryllium and boron depletion during
pre--main-sequence contraction. We calculate this limit by comparing
the nuclear luminosity $L_{\rm nuc}$ (i.e., the total rate of energy
generation due to p-p reactions in the star) to the energy release
rate due to contraction (eq.\ [\ref{eq:L-virial}]).  Note that stars
depleting lithium either burn it before p-p reactions become an
important source of energy (however, see \S~\ref{limitat}), or are not
massive enough to reach the main sequence.

The specific rate of heat production (\cite{cla83}) for the reactions
\nuc{p+p\to{}^2H+e^++\nu_e} and \nuc{^2H+p\to{}^3He+\gamma} is
\begin{equation}\label{eq:epsilon_pp}
\epsilon_{\rm pp} = 1.23\ee{6} \fscr X^2\rho T_6{}^{-2/3}
\exp\left(-\frac{33.8}{T_6{}^{1/3}}\right)\erg\secd^{-1}\gram^{-1},
\end{equation}
where $T_6\equiv T/10^6{\rm K}$, $\rho$ is in cgs units, $X$ is the
hydrogen mass fraction, and $\fscr$ is the screening correction factor
(Salpeter \& Van Horn 1969).  Further energy release from the PPI chain
will not occur until the abundance of \nuc{^3He} has increased to its
equilibrium value; the timescale for this is greater than $1\Gyr$ at
temperatures $\sim 5\times10^6\K$ (\cite{cla83}). Since the reaction
rate (eq.\ [\ref{eq:epsilon_pp}]) has such a steep temperature
dependence, most burning occurs near the center of the star.  Therefore,
we can approximate the screening factor $\fscr$ by its value at the
stellar center.  Moreover, if the star is fully convective, it is
described by an $n=3/2$ polytrope (see \S \ref{structure}) regardless of
the extent of nuclear burning. Under these conditions, the nuclear
luminosity is
\begin{equation}\label{eq:lnuc}
L_{\rm nuc}\equiv\int \epsilon_{\rm pp} dm \approx
4.72\ee{-2}T_{c6}^{0.36}\epsilon_{\rm pp}(\rho_c,T_c) M,
\end{equation}
where the prefactor in the above equation comes from integrating the
reaction rate over an $n=3/2$ polytrope and constructing a simple
analytic fit to the result.  Using equation (\ref{eq:lnuc}), we have
derived an analytic relation between the effective temperature along
the zero-age main sequence and stellar radius and surface gravity
(see Appendix \ref{s:ZAMSfits}).

When the nuclear luminosity is much less than the luminosity due to
gravitational contraction $L_c$ (eq.\ [\ref{eq:L-virial}]), the
assumptions leading to the contraction law (eq.\
[\ref{R-contraction}]) hold, and our depletion calculations are
reliable. We indicate in Figures \ref{fig:MassTemperature} and
\ref{fig:MassAge} the central temperatures and ages at which $L_{\rm
nuc}\approx 0.05 L_c$ and $L_{\rm nuc}\approx 0.50 L_c$ ({\em heavy
solid lines\/}).  For comparison, we also show ({\em triangles\/}) a
$1\Gyr$ isochrone (roughly, the main sequence) from CB97.  Our results
for light element depletion are only valid in the regions of the plots
below the aforementioned lines.  As one can see, lithium depletion is
unaffected by the hydrogen burning, while the beryllium results are
only valid for $M\gtrsim 0.1\msun$ (boron depletion will be discussed
in \S~\ref{s:B-ms-depl}).

Although p-p reactions can slow the contraction rate, they do not
alter the $n=3/2$ polytropic stellar structure.  However, for stars of
mass $M\gtrsim0.35\msun$ (\cite{cha97}) a radiative core forms during
contraction. Formation of a radiative core breaks the polytropic
structure, prevents mixing into the depletion region, and thus defines
the upper mass limit of applicability of our depletion calculations.
We estimate the central temperature at which convection ceases in the
core as follows.

The gas at the stellar center is convectively unstable whenever the
temperature gradient $d\ln T/d\ln P$ exceeds the adiabatic value
$1-1/\Gamma_2$ (see, for example, Hansen \& Kawaler 1994). In other words, the
maximum rate of energy generation {\em at the center of the star\/} that can
be balanced by radiative heat transport alone is
\begin{equation}\label{eq:MaxFlux}
\epsilon_{\rm rad}\equiv
	\left.\frac{\partial L(r)}{\partial m}\right|_{r=0}=
   	\frac{64\pi}{3} \frac{G\sigma_{\rm SB}}{\kappa} \frac{T_c{}^4}{P_c}
	\left(1-\frac{1}{\Gamma_2}\right),
\end{equation}
where $\kappa$ is the opacity.  The minimum rate of energy production
$\epsilon_{\rm rad}$ for which the stellar center is fully convective
is just equation (\ref{eq:MaxFlux}) with $\Gamma_2=5/3$.  In pre-main
sequence stars, the bulk of the energy is released by gravitational
contraction, with the specific energy production rate
\begin{equation}\label{eq:epsilon_c}
\epsilon_{\rm grav}=
	-\frac{P}{\rho\left(\Gamma_3-1\right)}
	\frac{d}{dt}\ln\left(\frac{P}{\rho^{\Gamma_1}}\right)=
	-\frac{3}{2}\frac{\kB\NA T}{\mueff}\frac{d\ln R}{dt};
\end{equation}
the total luminosity $L$ (equation [\ref{eq:L-virial}]) is just the
above rate integrated over the entire star.  As the star approaches
the main sequence, there is an additional small contribution
$\epsilon_{\rm pp}$ due to hydrogen burning (eq.\
[\ref{eq:epsilon_pp}]).  We equate the total energy generation rate at
the stellar center $\epsilon\equiv\epsilon_{\rm grav}+\epsilon_{\rm
pp}$ with $\epsilon_{\rm rad}$ and solve for $T_c$.  We compute the
opacity at the center from the OPAL tables (\cite{rog92}) and use
effective temperatures as a function of mass from CBP96 to compute the
contraction rate.  The resulting radiative/convective boundary curves
are plotted in Figures \ref{fig:MassTemperature} and \ref{fig:MassAge}
({\em heavy dot-dashed lines\/}).  The stars develop radiative cores
in the regions of the plots above the $\epsilon=\epsilon_{\rm rad}$
curves.  For \nuc{^7Li}, the mass, central temperature, and density at
which half-depletion is coincident with the formation of a radiative
core are $M_{1/2}\approx 0.5\msun$, $T_c\approx4.1\ee{6}\K$, and
$\rho_c\approx 4.8\gpcc$.  This is in excellent agreement with CB97,
who find that a radiative core forms in a $0.5\msun$ star at an age of
$10^7\yr$, when the lithium abundance is 0.52.  The corresponding
stellar parameters for which the formation of a radiative core is
coincident with \nuc{^7Li} depletion by a factor of 100 are
$M_{1/100}\approx0.44\msun$, $T_c\approx4.3\ee{6}\K$, and
$\rho_c\approx7.7\gpcc$.  For \nuc{^7Li}, our fully convective
depletion calculations are therefore valid for $M\lesssim
0.45-0.5\msun$, depending on the desired depletion level.

We should point out that the above results are quite sensitive to the
value of the opacity at the stellar center.  If we write the opacity
as $\kappa_\circ\rho^{\kappa_\rho}T^{\kappa_T}$ (with $\rho$ and $T$
in cgs units), we find that, at the central temperature where
\nuc{^7Li} is depleted by a factor of two,
$\kappa_\circ=5.66\ee{21}\cm^2\gram^{-1}$, $\kappa_\rho=0.347$, and
$\kappa_T=-3.166$.  The mass of the star that becomes radiative upon
depleting half of its \nuc{^7Li} content scales with $\kappa_\circ$
and $\Teff$ as $M_{1/2}\propto \kappa_\circ^{0.48} \Teff^{1.6}$.  The
sensitivity of observed lithium depletions to the interior opacity has
previously been considered for higher mass main-sequence stars by
\cite{swe90} and \cite{swe94}, who adjusted the interior opacities by
changing the metal content.  Although they were considering stars of
approximately solar mass, Swenson et al.\ (1990) found that an
increase of 37\% in the interior opacity increased the mass of a
Hyades star at which the lithium was half-depleted by $\sim 20\%$,
i.e., $M_{1/2}\propto\kappa_\circ^{0.5}$, similar to our estimated
dependence.

\section{Proton Capture Rates for Light Elements} \label{CaptureRates}

We use thermonuclear reaction rates in the form
\begin{equation}\label{eq:rr}
   \NA\langle\sigma v\rangle=S \fscr T_6{}^{-j} \exp
	\left(-\frac{a}{T_6{}^{1/3}}\right)\cm^3\secd^{-1}\gram^{-1},
\end{equation}
to approximate the rates of Caughlan \& Fowler (1988, hereafter CF88).
For non-resonant reactions $j=2/3$, while for reactions affected by
resonances the value of $j$ has to be adjusted (see below).  Here, $S$
does {\em not\/} represent the astrophysical S-factor, but is rather,
like $a$, a dimensionless parameter in the fit to the reaction rate.
Table \ref{t:reaction-rates} shows the values of $S$ and $a$ found by
fitting to the rates over the range of temperatures ($T_6<6$)
appropriate for this work.  Raimann (1993) recently incorporated new
experimental results at low energies ($\approx 11$--$13\keV$) to
update the rates for \nuc{^7Li}, adjusting the value of $S$ from
$6.4\ee{10}$ (CF88) to $7.2\ee{10}$.

There are two stable boron isotopes with abundance ratios in the local
ISM of \nuc{^{11}B}/\nuc{^{10}B}$=3.4^{+1.3}_{-0.6}$ (\cite{fed96}).
The \nuc{^{10}B(p,\alpha)\,^7Be} reaction is resonant, as an excited
state of the \nuc{^{11}C} nucleus exists just 10 keV above the
center-of-mass energy of the \nuc{p+\,^{10}B} system.  As a
non-resonant reaction, the center-of-mass energy would be at
$E_c=3.46T_6{}^{2/3}\keV$, and the reaction occurs (as we show later)
at $T_6\approx 6$, so that $E_c=11.4 \keV$, and thus this resonance
affects the cross section.  Rauscher \& Raimann (1996) recently
corrected the CF88 rates for this broad resonance, which enhances the
rate by at least a factor of 200 for the temperatures of interest.
For $T_6<6$, a reasonable fit to the reaction rates that includes the
resonance is to multiply the CF88 rates by $300(T_6/3)^{1/2}$.  The
entries for this reaction in Table \ref{t:reaction-rates} include the
resonant correction.

There are potential problems with the \nuc{^9Be\,(p,\alpha)\,^6Li}
rate as well, since an excited state of \nuc{^{10}B} ($E_x =
6.56\MeV$) lies just $25.7\keV$ below the energetic threshold. Current
knowledge (see \cite{ajz88}) holds that the angular momentum and
parity of this state is either $J^\pi=2^-$ or $J^\pi=4^-$, with $4^-$
thought to be the better guess.  The angular momentum and parity of
\nuc{^9Be} is $(3/2)^-$, while that of a proton is $(1/2)^+$.  Hence,
if this resonant state has $J^\pi=4^-$, the proton must overcome a
centrifugal barrier of angular momentum $\ell=2$, and the resonant
state will not contribute significantly to the reaction rate.  But if
instead $J^\pi=2^-$, the resonance can proceed via s-wave ($\ell=0$)
and the rate will be greatly enhanced over the CF88 rate (Brown 1997).

To estimate the effect such a resonance would have, we parameterize
the entrance channel by the dimensionless reduced width $\theta^2$
(\cite{cla83}).  Crudely speaking, $\theta^2$ measures the degree to
which the compound \nuc{p+^9Be} nucleus resembles the excited level of
\nuc{^{10}B}.  The cross section is then calculated using the
Breit-Wigner formula (see \cite{bro97} for details) with $\theta^2$ as
a free parameter.  A good fit to the thermally-averaged rate for
$J^\pi=2^-$ (accurate to within 1\% for $2<T_6<5$) is
\begin{equation}\label{eq:resfit}
   \NA\langle\sigma v\rangle \approx 3.53\ee{16}
      \theta^2 T_6{}^{-1.367} \exp\left(-\frac{105.33}
      {T_6{}^{1/3}}\right) \cm^3\gram^{-1}\secd^{-1};
\end{equation}
the total reaction rate is the sum of this equation and the standard
non-resonant rate.  For the temperatures of interest in low-mass
stars, $T_6\lesssim 6$, an s-wave resonance will dramatically boost
the rates if $\theta^2\gtrsim 0.01$. In this paper, in addition to
using CF88 rates for \nuc{^9Be}, we use equation (\ref{eq:resfit}) to
explore the effect of the resonance on beryllium depletion in
pre--main-sequence stars.

\section{The Depletion Equation}\label{formalism}

The central temperature needed for depletion can be roughly estimated
as follows.  Light elements are depleted when the nuclear destruction
time at the center of the star,
\begin{equation}\label{eq:tdest}
t_{\rm dest}\equiv\frac{1}{n_{\rm p}\langle\sigma v\rangle}
=0.689\left(\frac{M}{0.1\msun}\right)^2
\left(\frac{\mueff}{0.6}\right)^3
\frac{T_{c6}^{j-3}}{S\fscr}\exp\left(\frac{a}{T_{c6}^{1/3}}\right)\secd,
\end{equation}
where $n_{\rm p}$ is the proton density, becomes comparable to the
contraction time (i.e., the timescale for the change in central
temperature if degeneracy is mild),
\begin{equation}\label{eq:tcont}
t_{\rm cont}\equiv -\frac{R}{dR/dt}=
115\left(\frac{3000K}{\Teff}\right)^4
\left(\frac{0.1\msun}{M}\right)
\left(\frac{0.6}{\mueff}\right)^3
\left(\frac{T_c}{3\ee{6}}\right)^3 {\rm Myr}.
\end{equation}
The above equation is valid regardless of the degree of degeneracy so
long as the star is well described by an $n=3/2$ polytrope.  However,
for rough estimates, one can put
$\mueff=\mu\equiv(1/\mu_i+1/\mu_e)^{-1}$ in equations (\ref{eq:tdest})
and (\ref{eq:tcont}). Evaluating $t_{\rm dest}=t_{\rm cont}$ at the
stellar center gives an approximate condition for the central
temperature at the time of depletion,
\begin{equation}
\frac{a}{T_{c6}^{1/3}}=32.9+\ln\left(S\fscr\right)
-3\ln\left(\frac{M}{0.1\msun}\right)
-4\ln\left(\frac{\Teff}{3000\K}\right)
-6\ln\left(\frac{\mueff}{0.6}\right)
+(6-j)\ln T_{c6}.
\end{equation}
The destruction time is a very steep function of the central
temperature: once $T_c$ reaches a certain value, the entire abundance
of a light element is depleted on a timescale much shorter than
$t_{\rm cont}$. For example, a $0.1\msun$ star with $\Teff=3000\K$
burns \nuc{^7Li} when the central temperature $T_c\approx
3\ee{6}\K$. Under these conditions, the reaction rate is extremely
temperature sensitive ($\NA\langle\sigma v\rangle\propto T^{20}$) and
the central temperature necessary for elemental depletion depends only
weakly on mass.  The sharp dependence of the reaction rate on the
central temperature means that our calculations are extremely
insensitive to small uncertainties in the constitutive physics, such
as screening (\cite{bil97}).  For example, the central temperature at
depletion is proportional to $(S\fscr)^{1/20}$ and the time of
depletion $t\propto T_c^3\propto (S\fscr)^{3/20}$.  An uncertainty in
the reaction rate or the screening correction by a factor of 2 can
change the depletion time by no more than 10\%.  The exact equation
for elemental depletion (which we derive below) is just a fancy way of
writing $t_{\rm cont}=t_{\rm dest}$ that includes all the prefactors
that arise from integrating the destruction rate over the entire star,
and from the changing degeneracy (i.e., the decrease of $\mueff$)
during contraction.

At a given time $t$, consider a mass shell of coordinates $(m,m+dm)$
that contains $f(m,t)\,dm$ grams of a given light element isotope.
There are two contributions to the rate of change of $f(m,t)$:
(1) proton-capture reactions, and (2) convective transport from
isotope-rich regions to isotope-poor regions.  We ignore diffusion, as
it occurs over timescales too long to be of any relevance.
Symbolically,
\begin{equation}\label{eq:m1}
   \left(\ddt{f}\right)_m = \frac{f X \rho(m,t) }{m_H}
	\langle\sigma v\rangle (m,t) + \frac{\partial F}{\partial m},
\end{equation}
where $F$ is the net mass of the light isotope transported per second
by convection across a surface $r(m)$ at a time $t$, where $m_H$ is
the hydrogen mass.  The term $\partial F/\partial m$ vanishes upon
integration over the entire star.  Defining the mass-averaged value of
$f$ by the equation $M\bar{f} \equiv \int_0^M f\,dm$, we integrate
equation (\ref{eq:m1}) over the star to obtain
\begin{equation}\label{eq:m2}
   M \DDt{\bar{f}} = -\int_0^M \frac{Xf}{m_H}\rho\langle\sigma
	v\rangle\,dm .
\end{equation}
The timescale for a fluid element to traverse the radius is $t_{\rm
mix} \sim \ell/v$, where $\ell$ is the total length traveled.  For a
random walk process (\cite{bod65}), $\ell\sim R^2/\lambda_P$, where
$\lambda_P$ is the pressure scale height, which is of the order of the
mixing length.  The velocity may be estimated for the case of
efficient convection, giving $t_{\rm mix} \sim 10\mbox{--}100\yr$ in
the core.  Inserting a value of $\ell$ appropriate for the photosphere
gives $t_{\rm mix}\sim 10^3\yr$, which is still much shorter than the
evolutionary time scale of $10^6\mbox{--}10^8\yr$.  Therefore, the
star is well mixed and $f(m,t)$ can be replaced by $\bar{f}(t)$ in the
integral.  Then, upon changing variables to spatial coordinates in the
RHS of equation (\ref{eq:m2}), we have
\begin{equation}\label{eq:depl-rate}
\DDt{}\ln f = - \frac{4\pi X}{m_H M}\int_0^R \rho^2 \langle\sigma
v\rangle r^2\,dr.
\end{equation}
For simplicity of notation, we drop the bar over $f$.

Using the non-resonant form of the reaction rate (eq.\ [\ref{eq:rr}]
with $j=2/3$) and the definitions of central density and temperature
(eqs.\ [\ref{rho_c-polytrope}] and [\ref{T_c-polytrope}]), we write
the differential equation describing the change of mass fraction $f$
as
\begin{equation} 
   \frac{d}{dt}\ln f = -\frac{4\pi X}{\xi_1{}^3}
   \frac{\rho_c^2 R^3}{M}\frac{S}{\NA m_H}
   \left(\frac{\alpha}{a}\right)^2
   \int_0^{\xi_1} \fscr \xi^2 \Theta^{7/3} 
	\exp(-\alpha \Theta^{-1/3})\,d\xi \quad\secd^{-1},
\end{equation}
where $\alpha \equiv a T_{c6}{}^{-1/3}$. The burning rate is very
temperature-sensitive because the center of the star is much colder
than the Gamow energy. This sensitivity restricts the burning to the
central region of the star, where the temperature and density are
quadratic in radius.  Specifically, the peak of the rate integrand
lies at $\xi_{\rm peak} \approx 0.042 \left(55/\alpha \right)^2 +
0.0067 \left(55/\alpha \right)$.  For this same reason we approximate
the screening factor by its value at the center of the star.  We then
find an analytic approximation to the above equation that is accurate
to within $1\%$ for $\alpha > 45$ by using the first two terms of the
expansion $\Theta(\xi)=1-\xi^2/6+\ldots$ and taking the upper limit to
infinity.  This yields:
\begin{equation}\label{dlnfdt}
\frac{d}{dt} \ln f 
= - 0.18 
\left(\frac{X}{0.7}\right)
\left(\frac{0.6}{\mueff}\right)^{3}
\left(\frac{\msun}{M}\right)^{2}
S\fscr a^7 \alpha^{-17/2} 
\left(1-\frac{21}{2\alpha}\right) e^{-\alpha}\secd^{-1}.
\end{equation}
A similar development for resonant rates (with $j$ kept as a
parameter) yields
\begin{equation}\label{dlnfdt-boron}
\frac{d}{dt} \ln f 
= - 0.18
\left(\frac{X}{0.7}\right)
\left(\frac{0.6}{\mueff}\right)^{3}
\left(\frac{\msun}{M}\right)^{2}
S\fscr a^{9-3j} \alpha^{-(21/2-3j)} 
\left(1-\frac{27-9j}{2\alpha}\right) e^{-\alpha}\secd^{-1} .
\end{equation}
In order to integrate this equation, we need to know the dependence of
the central temperature parameter $\alpha$ on time, which is 
found by using the polytrope equation (\ref{T_c-polytrope}) and
the degeneracy relation (eq.\ [\ref{fhalf}]):
\begin{equation}\label{alpha}
\frac{\alpha}{a}=0.513\left(\frac{0.6}{\mueff}\right)^{1/3}
\left(\frac{\msun}{M}\right)^{1/3}\left(\frac{R}{\rsun}\right)^{1/3} 
=6.73\left(\frac{M}{\msun}\right)^{2/9}
\left(\frac{\mu_e F_{1/2}(\eta)}{t_6 T_{e3}^{4} }\right)^{2/9}.
\end{equation}
The value of $F_{1/2}(\eta)$ is obtained by solving the transcendental equation
(\ref{fhalf}).  Then we can use equation (\ref{alpha}) to integrate
the depletion equation (eq.\ [\ref{dlnfdt}] or eq.\
[\ref{dlnfdt-boron}]) numerically from $t=0$ and $\ln f=0$ to find the
element abundance as a function of time.  Results of the calculations
based on this formalism are presented in \S~\ref{results}.

In the region of the parameter space where degeneracy effects are not
dominant (i.e., to the right of the constant degeneracy line ({\it
heavy dotted line}) in Figure \ref{fig:MassTemperature}), the above
derivation can be considerably simplified, and the depletion equation
(either eq.\ [\ref{dlnfdt}] or [\ref{dlnfdt-boron}]) can be integrated
analytically (see \cite{bil97} for a simple version of this
development). During contraction, the central temperature initially
increases because the radius decreases (from the virial theorem for a
nondegenerate ideal gas $T_c\propto M/R$). For a nondegenerate gas,
$1/\mueff= 1/\mu_{\rm i}+1/\mu_{\rm e}$, but when degeneracy becomes
important $\mueff$ decreases, leading to a decrease in central
temperature.  Explicitly, we have
\begin{equation}
\frac{d}{dt}\ln f = \frac{d\ln f}{d\alpha}\frac{d\alpha}{dt} =
   \frac{d\ln f}{d\alpha} \left(\frac{\partial\alpha}{\partial
   R}\frac{dR}{dt} +
   \frac{\partial\alpha}{\partial\mueff}\frac{d\mueff}{dt}\right).
\end{equation}
When degeneracy is not important (roughly, for $M\gtrsim 0.2 \msun$), 
$\dot{\mu}_{\rm eff}$ is negligible compared to $\dot{R}$.  Thus, we
can rewrite the depletion equation (\ref{dlnfdt-boron}) in a simple form
\begin{eqnarray} \label{dt-depletion}
\frac{d}{d\alpha}\ln f &=& 6.23\ee{14} 
\left(\frac{X}{0.7}\right)\left(\frac{0.6}{\mueff}\right)^{6} 
\left(\frac{1000\K}{\Teff}\right)^4 \left(\frac{\msun}{M}\right)^3 \nonumber \\
	&& \times S \fscr a^{18-3j} 
\alpha^{-(41/2-3j)}
e^{-\alpha}\left(1-\frac{27-9j}{2\alpha}\right).
\end{eqnarray}
This equation may be integrated directly from the initial condition
$\alpha_0=\infty$ to $\alpha$: 
\begin{equation}\label{w-depletion} 
W \equiv \ln\frac{f_0}{f} = 6.23\ee{14} \left(\frac{X}{0.7}\right)
   \left(\frac{0.6}{\mueff}\right)^{6}\left(\frac{1000\K}{\Teff}\right)^4 
   \left(\frac{\msun}{M}\right)^3 S \fscr a^{18-3j} g(\alpha),
\end{equation}
where
\begin{equation}
g(\alpha)=\left[\alpha^{-(41/2-3j)} e^{-\alpha}
	-\frac{68-15j}{2}
	\Gamma\left(-(41/2-3j),\alpha\right)\right]
\end{equation}
and $\Gamma(x,\alpha)$ is the incomplete gamma function. For a given
$W$, equation (\ref{w-depletion}) can be solved for $\alpha(W)$, which
is essentially the central temperature required for the level of
depletion $W$. This is the method used in \cite{bil97} to compute
\nuc{^7Li} depletion.  We find that the two approaches (neglecting or
including $\dot{\mu}_{\rm eff}$) agree very well whenever the
approximate formulation, equation (\ref{dt-depletion}), is valid.

We note that the treatment of degeneracy and screening in equation
(\ref{w-depletion}) is approximate in the following sense.  In
addition to neglecting $\partial\alpha/\partial\mueff$ when changing
variables from $t$ to $\alpha$, we neglected the dependence of
$\mueff$ and $\fscr$ on $\alpha$ when integrating equation
(\ref{dt-depletion}).  This is justified because, when degeneracy is
mild, $\mueff$ and $\fscr$ vary very slowly with $\alpha$, while the
radius $R$ is a strong function of $\alpha$.  Furthermore, the
contraction rate of the star is $dR/dt\propto R^4$, so the star spends
most of the time at smaller radii, where the contraction rate is
small.  Thus, to approximate electron degeneracy and screening
corrections we just use the final values of $\mueff(\alpha,M)$ and
$\fscr(\alpha,M)$ in equation (\ref{w-depletion}) and solve the
resulting transcendental equation.  We find that the errors introduced
by this approximation, as compared to integrating equations
(\ref{dlnfdt}) or (\ref{dlnfdt-boron}) directly are never more than a
few percent.  For rough estimates that neglect effects of degeneracy
and screening, one can just set $1/\mueff=1/\mu_{\rm i}+1/\mu_{\rm
e}\equiv 1/\mu$ and $\fscr=1$.

\section{Discussion of the Results for Each Light Element}
\label{results}

We have presumed that the contracting pre--main-sequence star is well
described as a fully convective $n=3/2$ polytrope of arbitrary degeneracy with
no substantial internal energy generation. This limits our light element
depletion calculations in three ways:

\begin{enumerate}

\item {\bf Onset of Hydrogen Burning.} Hydrogen burning begins (see
\S\ \ref{coreconv}) before the star settles onto the zero-age main
sequence and thereby increases the time needed for the star to reach a
given radius, as compared to equation (\ref{R-contraction}). Since we
do not include this effect in our calculations, we limit our
discussion to those stars generating less than 5\% of their luminosity
by hydrogen burning.

\item {\bf Development of a Radiative Core.} Depletion at the stellar
surface is directly related to the central temperature only when the
star is fully convective. Further depletion will occur in the interior
parts of the star, but the surface abundance of a light element is
then sensitive to the temperature at the radiative/convective boundary
and not to the central temperature.  We stop calculating depletion
when the core of the star becomes radiative (see \S~\ref{coreconv}).

\item {\bf Coulomb Effects on the Equation of State.} Depending on the
particular element considered, Coulomb corrections to the equation of
state can be important at the time of depletion. These act to break
the polytropic nature of the star and increase the central temperature
relative to the polytropic value (see \S~\ref{limitat}).

\end{enumerate}

Figure \ref{fig:MassTemperature} displays our results for central
temperatures when the \nuc{^7Li} abundance varies from a half to a
hundredth of its initial value ({\em shaded region\/}).  We also plot
our central temperatures at one-half depletion of \nuc{^6Li} ({\em
light dotted line\/}) and \nuc{^9Be} ({\em light dashed line\/}).
(The boron isotopes are discussed in \S\ \ref{s:B-ms-depl}.)  The same
curves, expressed in terms of mass and age, are shown in Figure
\ref{fig:MassAge} and motivate the following discussions for each
element.

\subsection{Lithium Depletion}

The substantial amount of lithium\footnote{Throughout this section
references to lithium mean \nuc{^7Li}, as \nuc{^6Li} always depletes
first and is less abundant in the local ISM (Lemoine, Ferlet, \&
Vidal-Madjar 1995).} observations already makes the case that lithium
abundances are a reliable indicator of stellar ages and masses
(Rebolo, Mart\'{\i}n, \& Magazz\`u 1992; \cite{mar94}; \cite{bas96};
\cite{reb96}; \cite{opp97}).  In \S~\ref{alphaper} we present an
extended discussion of the observations of the cluster Alpha Persei
and derive a minimum age of $60\Myr$. Since this same technique can be
applied to clusters in the age range of $10$--$200\Myr$, we
extensively discuss the application of our results to the lithium
observations.

Previously (in \cite{bil97}), we discussed the lithium depletion
results and provided convenient fitting formula. In addition to
providing slightly improved formulae here, we also discuss in detail
the limitations of our work at the very low-mass end. The intersection
of the lithium depletion with the onset of a radiative core (see
Figures \ref{fig:MassTemperature} and \ref{fig:MassAge}) confines our
work to $M\lesssim0.46\msun$ (for this mass, the ending of convection
in the core is concurrent with depletion of Li to 1\% of its original
abundance).  Our method has the advantage of leaving $\Teff$
independent of stellar mass, but the disadvantage of not including
Coulomb effects, which limits us to masses $M\gtrsim\mcoul$, or, for
typical values of $\Teff$, to stars younger than $\sim 200-300\Myr$.

\subsubsection{Comparison to Previous Theoretical Work and
Limitations at Low Masses}
\label{limitat}

Based on the results presented in \cite{bil97}, and using a fiducial
effective temperature relation, $\Teff\propto M^{1/7}$, we obtain the
central temperature at the time of one-half lithium depletion,
$T_c\approx 3.1 \times 10^6 \K (M/0.1 \msun)^{1/6}$.  From the
polytropic relation (eq.\ [\ref{rho_c-polytrope}]), the central
density at this time is $\rho_c\approx 62 \gram\cm^{-3} (0.1
\msun/M)^{3/2}$.  From $T_c$ and $\rho_c$, we assess how important
Coulomb corrections to the equation of state will be at the center of
the star when the burning occurs. The plasma parameter at the time of
one-half depletion is
\begin{equation}
\Gamma(1/2 {\rm\ ^7{Li}\ depletion})\approx {e^2\over k_BT} 
\left({4\pi\rho\over 3 m_H}\right)^{1/3}
\approx 0.3 \left(0.1 \msun\over M\right)^{2/3}.
\end{equation}
Accurate evaluations of the equation of state in this regime are known
(see Saumon et al.\ 1996). However, we will roughly estimate the size
of the Coulomb correction using the Coulomb pressure for $\Gamma \ll
1$ (Clayton 1983), $P_{\rm C}=-e^3/3(\pi/k_B T)^{1/2} (1.8
\rho/m_H)^{3/2}$, which then gives
\begin{equation}
{P_{\rm C}\over P}\approx -6.8\times 10^{-2}
\left(0.1 \msun\over M\right), 
\end{equation}
at the time of one-half depletion. Since the Coulomb pressure is
negative, the central temperature is higher than its polytropic value.

\label{MaximumTemperature}	%leave the label with following
Our neglect of the Coulomb correction places a lower limit on the mass
of a star for which we can reliably compute the light element
depletion.  It is easy to show that, for normal composition ($X=0.7$),
the central temperature as a function of degeneracy is
\begin{equation}
T_c=5.63\ee{6} \left(\frac{M}{0.075\msun}\right)^{4/3}
\left(\mueff F_{1/2}^{1/3}(\eta)\right)^{2} {\rm K}.
\end{equation}
As the star contracts, the central temperature rises until the degree
of degeneracy reaches a critical value ($F_{1/2}(\eta)\approx 4-6$),
after which the central temperature begins to decline.  The maximum
value of $T_c$ is
\begin{equation}\label{eq:MaximumTemperature}
T_{c,{\rm max}}=3.43\ee{6}\left(\frac{M}{0.075\msun}\right)^{4/3} {\rm K}.
\end{equation}
We note that equation (\ref{eq:MaximumTemperature}) fully incorporates
the effects of partial electron degeneracy, while a similar relation
for $T_{c,{\rm max}}$ using an approximate treatment of degeneracy is
derived in Stevenson (1991) and Burrows \& Liebert (1993).  The
maximum temperature as a function of mass (for $M<0.1\msun$) is
denoted with a thick dashed line in Figure \ref{fig:MassTemperature}.
The intersection of this curve with the lithium depletion line defines
the lower mass limit of validity of our lithium depletion
calculations. 

For high-mass stars that deplete lithium long before the central
temperature reaches its peak, our neglect of Coulomb corrections
slightly delays the time of depletion. This is because the Coulomb
corrections allow the star to reach a given $T_c$ at an earlier time
and thereby deplete earlier. Roughly, since the time to reach a given
central temperature scales as $T_c^3 $, we expect a 2\% error in the
central temperature to give a 6\% error in time. However, the stars
that deplete lithium after (or while) contracting past the maximum
temperature point always have $t_{\rm dest}\gg t_{\rm cont}$.  Small
corrections to the equation of state then have a large effect on the
depletion time.  Thus, our results are only reliable for stars that
deplete lithium before reaching the maximum central temperature point,
or that are younger than
\begin{equation}
t=235 \left(\frac{M}{0.075\msun}\right)^3
\left(\frac{2800K}{\Teff}\right)^4 \Myr,
\end{equation}
which limits us to masses $M\gtrsim 0.072\msun$ for lithium depletion
by a factor of 100.  The exact value of the low-mass cutoff is
sensitive to the effective temperature, and Figure \ref{fig:hump}
displays the dependence of the cutoff mass on $\Teff$ and the level of
depletion.  For a given level of depletion, our calculations are valid
only to the right of the corresponding curve in Figure
\ref{fig:hump}. For masses lower than the above cutoff, one needs to
include all of the relevant corrections to the equation of state in
order to reliably predict lithium depletion at low levels and we refer
the reader to the papers of CB97 and DM94 for accurate calculations in
this regime.

In order to simplify the use of our results in evaluating observations
and comparing to detailed calculations, we have constructed simple
analytic fits to the solutions of equations (\ref{dlnfdt}) and
(\ref{dlnfdt-boron}).  The fits for age and central temperature of a
star at a certain stage of \nuc{^7Li} depletion (good for
$\mcoul\lesssim M\lesssim 0.44\msun$ for \nuc{^7Li} depletion by a
factor of 100; the exact range is outlined in Figures
\ref{fig:MassTemperature} and \ref{fig:hump}) are
\begin{eqnarray}\label{eq:Li7AgeFit}
t\left(^7{\rm Li}\right) &=& 50.6
   \left(\frac{0.1\msun}{M}\right)^{0.707}
   \left(\frac{3000\K}{\Teff}\right)^{3.516}
   \left(\frac{\mu_\circ}{\mu}\right)^{2.08}
   \left(\frac{W}{\ln2}\right)^{0.121}\\ \nonumber &&
   \times\exp\left[ 0.0951 \left(\frac{0.1\msun}{M}\right)^{5.02}
   \left(\frac{\Teff}{3000\K}\right)^{0.599}
   \left(\frac{\mu_\circ}{\mu}\right)^{8.577}
   \left(\frac{W}{\ln2}\right)^{0.148} \right] \Myr
\end{eqnarray}
and
\begin{eqnarray}\label{eq:Li7TcFit}
T_{c}\left(^7{\rm Li}\right) &=&
   3.06\ee{6}\left(\frac{M}{0.1\msun}\right)^{0.149}
   \left(\frac{\Teff}{3000\K}\right)^{0.158}
   \left(\frac{\mu}{\mu_\circ}\right)^{0.341}
   \left(\frac{W}{\ln2}\right)^{0.039} \\ \nonumber &&
   \times\exp\left[ -0.0149 \left(\frac{0.1\msun}{M}\right)^{5.54}
   \left(\frac{\Teff}{3000\K}\right)^{0.782}
   \left(\frac{\mu_\circ}{\mu}\right)^{9.37}
   \left(\frac{W}{\ln2}\right)^{0.197} \right]\K,
\end{eqnarray}
where $\mu_\circ=4/(3+5X)$ for $X=0.7$ and $W=\ln f_\circ/\ln f$.  The
accuracy of these fits (as well as all the other fits presented below)
is better than 1\% for the age and 0.3\% for the central
temperature. The corresponding fits for \nuc{^6Li} are
\begin{eqnarray}\label{eq:Li6AgeFit}
t\left(^6{\rm Li}\right) &=& 30.1
   \left(\frac{0.1\msun}{M}\right)^{0.72}
   \left(\frac{3000\K}{\Teff}\right)^{3.532}
   \left(\frac{\mu_\circ}{\mu}\right)^{2.179}
   \left(\frac{W}{\ln2}\right)^{0.117} \\ \nonumber &&
   \times\exp\left[ 0.0466 \left(\frac{0.1\msun}{M}\right)^{5.088}
   \left(\frac{\Teff}{3000\K}\right)^{0.575}
   \left(\frac{\mu_\circ}{\mu}\right)^{8.602}
   \left(\frac{W}{\ln2}\right)^{0.142} \right] \Myr
\end{eqnarray}
and
\begin{eqnarray}\label{eq:Li6TcFit}
T_{c}\left(^6{\rm Li}\right) &=& 2.58\ee{6}
   \left(\frac{M}{0.1\msun}\right)^{0.144}
   \left(\frac{\Teff}{3000K}\right)^{0.15}
   \left(\frac{\mu}{\mu_\circ}\right)^{0.344}
   \left(\frac{W}{\ln2}\right)^{0.037} \\ \nonumber &&
   \times\exp\left[-0.006159 \left(\frac{0.1\msun}{M}\right)^{5.7}
   \left(\frac{\Teff}{3000\K}\right)^{0.755}
   \left(\frac{\mu_\circ}{\mu}\right)^{9.55}
   \left(\frac{W}{\ln2}\right)^{0.19} \right]\K.
\end{eqnarray}
In the above equations the exponential factor accounts for the effects
of degeneracy; for $M\ge 0.2\,\msun$ it can be set to 1 without
sacrificing the accuracy of the fit.

In our previous paper (\cite{bil97}), we compared our calculations for
central temperature and age as functions of lithium depletion to those
of CBP96.  We found that the discrepancies between our results (using
their effective temperatures) and theirs increased with the amount of
lithium depletion. This difference has now been resolved. In the
calculation of CBP96, the timestep did not scale with the rapidly
decreasing lithium depletion timescale; as a result, when $t_{\rm
dest}$ became much less than the Kelvin-Helmholtz timescale, the
change in lithium abundance was no longer accurately computed.  With
the timestep adjusted to resolve $t_{\rm dest}$, their code produces
central temperatures (ages) consistent with ours to within 2\% (10\%)
for $0.09<(M/\msun)<0.2$, and to within 3\% (15\%) for
$0.2<(M/\msun)<0.35$ (Baraffe 1997).  Remaining discrepancies are now
of comparable size to the differences between our results and those of
DM94 (using the effective temperatures of DM94), and give us renewed
confidence in our semi-analytic calculations.  We also note that,
since the time of elemental depletion is rather sensitive to the
effective temperature ($t\propto\Teff^{-3.5}$), the typical
observational errors of $\sim 5\%$ in $\Teff$ lead to an uncertainty
of $\sim 15\%$ in the age determination.  On the other hand, the
corresponding central temperature is rather insensitive to the errors
in $\Teff$ (\cite{bil97}).

\subsubsection{Comparison to Observations: Alpha Persei}
\label{alphaper}

As an example of how our depletion calculations allow us to place
constraints on ages of young clusters (first carried out in the
Pleiades by Basri et al.\ 1996 and \cite{bil97}) we apply our work to
the recently reported non-detection of lithium in faint objects in the
$\alpha$ Per open cluster (Zapatero Osorio et al.\ 1996).  We use
their observed luminosities to infer the minimum age of the cluster,
independent from the uncertainties in the effective temperature scale.
From this minimum age and the observed effective temperatures we
constrain the stellar masses.

As described in \cite{bil97}, if lithium is undetected to some limit
$W=\ln(f_\circ/f)$ in a star with $-3.50 < \log(L/L_\odot) < -2.32$ then
the star must be older than
\begin{equation}\label{eq:tmin}
t_{\rm min} = 54.7 \left(\frac{10^{-2.5}\Lsun}{L}\right)^{0.922}
   \left(\frac{\mu_\circ}{\mu}\right)^{2.52} 
   \left(\frac{W}{\ln 2}\right)^{0.0769} \Myr.
\end{equation}
This minimum age does not depend on the effective temperature; rather,
the minimum arises because electron-degeneracy pressure becomes
important as low-mass stars contract (\cite{bil97}).  Equation
(\ref{eq:tmin}) therefore describes an absolute minimum age for a
lithium depleted star in the proper luminosity range, independent of
the uncertainties in the effective temperature scale. It is the
dimmest star with no lithium that sets the most restrictive bound on
the minimum age.  Applying equation (\ref{eq:tmin}) to
AP~279\footnote{The latest spectral type $\alpha$ Per member
considered by Zapatero Osorio et al.\ (1996) is AP~J0323+4853.
However, it is slightly brighter ($\log(L/L_\odot)=-2.46\pm 0.06$)
than AP~279, and hence yields a less stringent minimum age.}
($\log(L/\Lsun)=-2.49\pm0.05$), we find that the cluster, if coeval,
must be older than $61\pm 7\Myr$, where the error comes from the
uncertainty in the luminosity.  The most recent age determination for
this cluster from main-sequence turnoff by Meynet et al.\ (1993) gave
53~Myr for moderate convective overshoot, and the conventional turnoff
age without convective overshoot is 51~Myr (\cite{mer81}).  As with
lithium dating of young stars in the Pleiades (Basri et al.\ 1996), we
find that the nuclear age determination using lithium observations is
consistent with traditional upper main-sequence fitting only if
convective overshoot plays a significant role in the evolution of
massive stars.  The minimum age derived above implies that $\alpha$
Per is likely to be older than what has been previously inferred from
main-sequence turnoff, even with moderate convective overshoot, in
agreement with the conclusion of Zapatero Osorio et al.\ (1996).
However, because of the uncertainty in the luminosity of AP~279, and
if we allow a range of a few Myr in the ages of individual cluster
stars, we cannot rule out the possibility that the lithium nuclear age
and the conventional main-sequence turnoff age are consistent.  If dimmer
lithium-depleted stars are found in $\alpha$ Per, this consistency
will be harder to maintain (\cite{bas98}).

Lithium ages of $\alpha$~Per and Pleiades are sensitive to the
luminosities of their dimmest lithium-depleted members, $t_{\rm
Li}\propto L^{-0.9}$ (eq.\ [\ref{eq:tmin}]), and, hence, to the
assumed distance to the clusters.  However, since the age determined
from the main-sequence turnoff scales as $t_{\rm MS}\propto L^{-0.6}$
(\cite{mey93}), the {\it difference} between $t_{\rm Li}$ and $t_{\rm
MS}$ is a weak function of the assumed distance.  For example, the
Pleiades would need to be at least a factor of $2$ more distant for
the lithium age to coincide with the upper main-sequence age with no
convective overshoot.

Given the inferred minimum age, we use the observed luminosity and lithium
abundance to set a lower bound on each star's mass.  Combining
equations (\ref{eq:L-virial}) and (\ref{R-contraction}) yields the
mass of a star that has a certain luminosity and age,
\begin{equation}\label{eq:mass-age}
   \frac{M}{\msun} = 0.122 \left(\frac{L}{10^{-2.5}\Lsun}\right)^{3/4}
   \left(\frac{t}{100\Myr}\right)^{1/2} \left(\frac{3000\K}{\Teff}\right).
\end{equation}
Substituting the value of the minimum age into this equation gives the
minimum mass for each cluster member that is fully convective and has
not yet arrived on the main sequence.  We illustrate this technique in
Figure \ref{fig:aper} for the stars AP279, AP272, and AP284. The
shaded regions for each star represent the allowed values of mass and
age given their level of lithium depletion.  If we presume that they
are all of an age equal to or larger than our minimum estimate, then
we find the following minimum masses: AP143, $0.30\msun$; AP 296,
$0.25\msun$; AP284, $0.18\msun$; AP268, $0.11\msun$; AP 272,
$0.12\msun$; AP279, $0.09\msun$; and AP~J0323+4853\footnote{It has
recently been reported (\cite{mar97}) that the rotation period of
AP~J0323+4853 is $7.6\,{\rm hr}$.  Using the formula for the radius
(eq.\ [\ref{R-contraction}]), we estimate the breakup spin period
($P_b\approx 2 \pi (R^3/GM)^{1/2}$) for the best guess at minimum age
($61.2\Myr$) and mass ($0.09\msun$) to be $\approx 0.13 \,{\rm hr}$.
Rotation is therefore unlikely to cause any deviations of the
depletion time from our estimates (e.g., by decreasing the central
temperature).}, $0.09\msun$. It is important to keep in mind that the
minimum mass determination, unlike the minimum age determination,
depends on the effective temperature scale used, and so is subject to
systematic errors.

\subsection{Beryllium Depletion}

Since our calculations require the star to be contracting and fully
convective, or below the nuclear burning and radiative core lines in Figure
\ref{fig:MassTemperature}, it is clear that our theory for \nuc{^9Be}
(standard rate) applies to stars with masses in the range $0.1$--$0.3\,\msun$.
All of these objects clearly deplete beryllium prior to reaching the main
sequence, and for this mass range the age of the star at a given level of
depletion is
\begin{eqnarray}\label{eq:Be9AgeFit}
t\left({\rm ^9Be}\right) &=& 99.6 \left(\frac{0.1\msun}{M}\right)^{0.711}
   \left(\frac{3000\K}{\Teff}\right)^{3.555}
   \left(\frac{\mu_\circ}{\mu}\right)^{2.172}
   \left(\frac{W}{\ln2}\right)^{0.111}\\ \nonumber
&& \times\exp\left[ 0.3365 \left(\frac{0.1\msun}{M}\right)^{3.60}
      \left(\frac{\Teff}{3000\K}\right)^{0.424}
      \left(\frac{\mu_\circ}{\mu}\right)^{6.582}
      \left(\frac{W}{\ln2}\right)^{0.098}
   \right] \Myr,
\end{eqnarray}
while the central temperature is
\begin{eqnarray}\label{eq:Be9TcFit}
T_{c}\left({\rm ^9Be}\right) &=& 3.92\ee{6}
   \left(\frac{M}{0.1\msun}\right)^{0.129}
   \left(\frac{\Teff}{3000\K}\right)^{0.146}
   \left(\frac{\mu}{\mu_\circ}\right)^{0.313}
   \left(\frac{W}{\ln2}\right)^{0.036}\\ \nonumber
&& \times\exp\left[-0.057261 \left(\frac{0.1\msun}{M}\right)^{3.408}
      \left(\frac{\Teff}{3000\K}\right)^{0.521}
      \left(\frac{\mu_\circ}{\mu}\right)^{6.228}
      \left(\frac{W}{\ln2}\right)^{0.134}
   \right]\K.
\end{eqnarray}
Stars with $M\lesssim 0.1\msun$ deplete beryllium while approaching
the main sequence or residing on it.  For these objects, we refer the
reader to the results of Nelson et al.\ (1993) and CB97.  For masses
$0.1\msun<M<0.3\msun$, our central temperatures (ages) differ by no
more than 2\% (13\%) from those of CB97.  Our comparisons are for
abundances of more than 0.5 of the initial abundance, on account of
the aforementioned error in the calculations of CB97.

If the reaction \nuc{^9Be\,(p,\alpha)\,^6Li} does proceed via a sub-threshold
resonance (see \S~\ref{CaptureRates}), then parameterizing the cross section
by the reduced width (see eq.\ [\ref{eq:resfit}]) implies that the age and
central temperature at a given depletion level are
\begin{eqnarray}\label{eq:Be9resAgeFit}
\lefteqn{t\left({\rm ^9Be}\right) =
   100.8\left(\frac{0.1\msun}{M}\right)^{0.699}
   \left(\frac{3000\K}{\Teff}\right)^{3.547}
   \left(\frac{\mu_\circ}{\mu}\right)^{2.142}
   \left(\frac{W}{\ln2}\right)^{0.113}
   \left(\frac{0.1}{\theta}\right)^{0.227}}\\ \nonumber
&& \times\exp\left[ 0.3435 \left(\frac{0.1\msun}{M}\right)^{3.612}
      \left(\frac{\Teff}{3000\K}\right)^{0.46}
      \left(\frac{\mu_\circ}{\mu}\right)^{6.8}
      \left(\frac{W}{\ln2}\right)^{0.105}
      \left(\frac{0.1}{\theta}\right)^{0.217}
   \right] \Myr
\end{eqnarray}
and
\begin{eqnarray}\label{eq:Be9resTcFit}
\lefteqn{T_{c}\left({\rm ^9Be}\right) = 
   4.19\ee{6}\left(\frac{M}{0.1\msun}\right)^{0.129}
   \left(\frac{\Teff}{3000\K}\right)^{0.149}
   \left(\frac{\mu}{\mu_\circ}\right)^{0.314}
   \left(\frac{W}{\ln2}\right)^{0.037}
   \left(\frac{0.1}{\theta}\right)^{0.075}}\\ \nonumber
&& \times\exp\left[-0.0653
      \left(\frac{0.1\msun}{M}\right)^{3.101}
      \left(\frac{\Teff}{3000\K}\right)^{0.542}
      \left(\frac{\mu_\circ}{\mu}\right)^{6.372}
      \left(\frac{W}{\ln2}\right)^{0.132}
      \left(\frac{0.1}{\theta}\right)^{0.266}
   \right]\K.
\end{eqnarray}
For example, a star of mass $0.2\,\msun$ at half-depletion will be
22\% younger if the resonance exists and $\theta = 0.3$.

\subsection{Boron Depletion}
\label{s:B-ms-depl}

Unlike lithium and beryllium, the higher Coulomb barrier of boron
means that it is depleted when either hydrogen burning or radiative
heat transport is important.  In Figure \ref{fig:boron}, we show the
central temperature at the time when one-half of \nuc{^{10}B} ({\em
thin solid line\/}) and \nuc{^{11}B} ({\em thin dotted line\/}) are
depleted, {\em assuming} that the stars are fully convective and not
burning hydrogen.  As one can see, this assumption holds only for a
very narrow range of masses for \nuc{^{10}B}, and does not hold for
\nuc{^{11}B} at all.  Thus we can only make qualitative conclusions
about boron depletion during pre--main-sequence contraction.  Only
Nelson et al.\ (1993) and CB97 have tabulated the boron depletion.
Neither of these authors displayed results for the less abundant
isotope \nuc{^{10}B}. If the resonance were not present in the
\nuc{p+{}^{10}B} reaction (see \S~\ref{CaptureRates}), then
\nuc{^{10}B} would deplete after \nuc{^{11}B} and their boron results
would have been in error.  However, Figure \ref{fig:boron} clearly
shows that the resonance-corrected rate for this reaction is large
enough so as to make it deplete before \nuc{^{11}B}, in which case the
aforementioned calculations are indeed valid.

Our approximations are, however, applicable to stars in a certain mass
range that deplete boron on the main sequence.  While on the main
sequence, the central temperature of a star is essentially fixed.
Moreover, stars with $M\lesssim0.3\msun$ are fully convective on the
main sequence.  For such stars the formalism developed in
\S~\ref{formalism} is fully applicable, and, in particular, equations
(\ref{dlnfdt}) and (\ref{dlnfdt-boron}) with $\alpha={\rm const}$
imply that the abundances of \nuc{^{10}B} and \nuc{^{11}B} will decay
exponentially, at rates determined by the star's central density and
temperature. Exponential decay of \nuc{^{11}B} is consistent with its
abundances at 1 and 10 Gyr in Table 2 of CB97; for stars of mass
$0.085\msun\lesssim M\lesssim 0.13\msun$, this depletion happens on
the main sequence before the Hubble time. Moreover, the depletion time
scale is an extremely strong function of mass within this mass range.

It is clear that boron abundance in low-mass ($M\lesssim 0.13\msun$)
main-sequence stars may provide a valuable constraint on their masses
and ages, if boron can be observed. However, the strong dependence of
depletion on the central temperature, as well as the high central
density of these stars, necessitate sophisticated calculations like
those of CB97 to accurately relate depletion time scales to stellar
masses.

\section{Conclusions}\label{Conclusions}

The ability of stars to burn lithium, beryllium, and boron prior to
reaching the main sequence has been recognized for quite some
time. However, the observations of such depletion are just now
becoming accessible for pre--main-sequence stars in nearby clusters.
This has motivated our semi-analytic work on light element depletion
for low-mass ($M\lesssim0.5 \msun$) pre--main-sequence stars and the
application of our results to observations of lithium-depleted stars.
Lithium observations are proving to be an excellent age indicator in
open clusters (see \cite{bil97} for a more extensive discussion) and
our new minimium age determination for $\alpha$ Per (60 Myr) is in
reasonable agreement with that found by upper main-sequence modeling
with mild convective overshoot.  However, if dimmer lithium-depleted
stars are found in $\alpha$~Per, more overshoot may be necessary in
order to reconcile the lithium nuclear age with the upper
main-sequence turnoff one (\cite{bas98}).  In addition, the difference
between the nuclear and upper main-sequence turnoff ages is rather
insensitive to the distance to the cluster.  We encourage observers to
map out the lithium depletion region for pre--main-sequence stars in
clusters of ages 10--200 Myr. This will provide an important check on
the mass dependence of convective overshoot in the massive stars near
the turnoff in these clusters.  The main results of this paper are the
fitting formulae for the age and central temperature as a function of
mass, effective temperature, composition, and depletion level for
\nuc{^7Li} (eq.\ [\ref{eq:Li7AgeFit}] and [\ref{eq:Li7TcFit}]),
\nuc{^6Li} (eq.\ [\ref{eq:Li6AgeFit}] and [\ref{eq:Li6TcFit}]), and
\nuc{^9Be} (eq.\ [\ref{eq:Be9AgeFit}] and [\ref{eq:Be9TcFit}]). These
relations are robust and useful for evaluating observations and for
comparing to detailed numerical calculations.  Moreover, our results
allow using the observed $\Teff$ directly, instead of first placing a
star on a theoretical evolutionary track, as well as exploring the
effects of the different observational $\Teff$ scales.

Lithium dating of open clusters will most certainly help answer the
question of how massive a star can become a white dwarf (WD). Ever
since the white dwarf survey of young clusters of Romanishin \& Angel
(1980), there have been spectroscopic followups on the WD candidates
(Reimers \& Koester 1988). In addition to measuring the masses of the
WD's, the ultimate goal of these efforts is to learn the range of
stellar masses that can become isolated WD's. Crucial to this program
is the ability to subtract the white dwarf cooling age from the
cluster age so as to get the white dwarf birthday and thereby the mass
of the progenitor star at that time.  The most recent work of this
nature is in NGC 2516 (Koester \& Reimers 1993), which has an age
infered from the upper main-sequence turnoff of 140 Myr (Meynet et
al.\ 1993), giving an inferred mass of the progenitor star of around
$6 \, \msun$ (see also Reid 1996 for detailed discussion). Lithium
dating of this (and other) young clusters would enhance our confidence
in the extrapolation required to obtain the WD progenitor mass.

In the cold atmosphere of a low-mass star, the typically observed
absorption lines of beryllium, 3130 and 3131\AA, and boron, 2497 and
2498\AA\ (\cite{boe96}), lie in a part of the continuum where there
are nearly no photons. Observations of pre--main-sequence depletion
similar to those carried out for lithium thus await more sensitive
instruments. There is, however, one intriguing possibility for
diagnosing the elemental abundance of the surfaces of the low-mass
main-sequence stars that might not have depleted beryllium or
boron. Some of these low-mass objects are known to reside in accreting
binaries with white dwarfs, especially those with orbital periods
below 2 hours (Warner 1995). Most of these binaries are dwarf novae,
in which the matter leaving the low-mass star due to Roche lobe
overflow accumulates in the disk for some time before reaching a
thermal instability strong enough to dump the matter onto the white
dwarf. There have been recent HST observations of the white dwarf
photosphere after these accretion events (Sion 1995; Sion et al.\
1997; Cheng et al.\ 1997) which show that the WD is heated by
accretion. Since the matter at the photosphere is actually from the
low-mass star, there may be a chance of seeing some features from the
beryllium and boron that are now residing in the WD photosphere, where
the continuum in the blue/UV is much higher.  This approach is
complicated by the fact that nearly all binary evolution calculations
suggest that low-mass objects in systems with such short periods have
experienced substantial mass transfer. This means that the progenitors
of these low-mass stars are likely to have been more massive in the
past, and may have depleted beryllium and boron in their cores.

\acknowledgements 

We thank Gibor Basri, Lynne Hillenbrand, Eduardo Mart\'{\i}n, and
Maria Rosa Zapatero Osorio for keeping us up to date on the
observational situation regarding lithium depletion and comments on
the manuscript, Isabelle Baraffe for numerous comparisons with our
results, Gerhard Raimann and Thomas Rauscher for communications about
proton capture on boron, and the anonymous referee for his/her helpful
comments. G.~U. thanks the Fannie and John Hertz Foundation for
fellowship support.  C.~D.~M. was supported by an NSF Graduate
Research Fellowship, and E.~F.~B. was supported by a NASA GSRP
Graduate Fellowship under grant NGT-51662.  L.~B. acknowledges support
as an Alfred P. Sloan Foundation fellow.
 
\newpage

\appendix

\section{Fitting Formula for the Zero-Age Main Sequence}
\label{s:ZAMSfits}

Using the formalism presented in \S~\ref{coreconv}, we present a
convenient fitting formula for the zero-age mainsequence (ZAMS)
radius as a function of mass and effective temperature,
\begin{equation}
   \frac{R}{\rsun} = 0.129 \left(\frac{M}{0.1\msun}\right)^{0.766}
   \left(\frac{\Teff}{3000\K}\right)^{-0.397} \exp\left[-0.192
   \left(\frac{M}{0.1\msun}\right)^{-5.73}
   \left(\frac{\Teff}{3000\K}\right)^{1.48} \right].
\end{equation}
This fit is good for $0.1<M/\msun<0.3$ and $2500\K<\Teff<3500\K$.  As
a test of our approximations, we compare our radii to 1~Gyr isochrones
from CB97 and DM94.  We take the effective temperatures from the
isochrone for each mass.  Our radii deviate by no more than 2\%
($0.15<M/\msun<0.3$) from Table 2 of CB97, and the maximum deviation
for $0.1<M/\msun<0.3$ is 10\%.  The increasing deviation at lower
masses is due to the Coulomb correction to the equation of state.
From Figure \ref{fig:MassTemperature}, the 1~Gyr isochrone of CB97
lies slightly above the maximum temperature reachable with our
approximations (i.e., with Coulomb effects neglected).  Our radii
deviate no more than 8\% (7\%) from those inferred from Table 5 (6) of
DM94.

In Figure \ref{fig:gTeff}, we plot the surface gravity $\log\,g$ as a
function of $\Teff$ for $M/\msun = 0.12$, 0.16, 0.20, 0.24, and 0.30.
Unfortunately, there is only one binary with stars in the mass range
of our fit.  This is the binary CM~Draconis (\cite{lac77}), which has
a primary (secondary) of mass $(0.237\pm0.011)\msun$
[$(0.207\pm0.008)\msun$] and effective temperature of $(3150\pm100)\K$
(same for the secondary).  Using these parameters, we infer a radius
of $0.246\rsun$ ($0.221\rsun$) and a surface gravity $\log\,g=5.03$
($\log\,g=5.07$); the values from Lacy (1977) are
$(0.252\pm0.008)\rsun$ [$(0.235\pm0.007)\rsun$] and $5.01\pm0.05$
(same for the secondary).

Similar fits were recently presented by Tout et al.\ (1996) for the ZAMS
radii as functions of mass and metallicity.  Our fits have the advantage
of being in terms of mass and effective temperature, and so are
independent of the model assumptions that are incorporated in mapping
the metallicity to the effective temperature.

\begin{deluxetable}{llllc}
%\tablewidth{250pt}
\tablecaption{%
Proton Capture Rate Parameters
\label{t:reaction-rates}}
\tablehead{%
%------------------------------------------------------------------------
\colhead{Reaction} & \colhead{$S$} & \colhead{$a$} & \colhead{$j$} & \colhead{References}
}
%------------------------------------------------------------------------
\startdata
\nuc{^6Li(p,\,^3He)\,^4He} 	& $3.75\ee{12}$ 	& \phn$84.13$	& $2/3$	& 1	\nl
\nuc{^7Li(p,\alpha)\,^4He} 	& $7.20\ee{10}$ 	& \phn$84.72$	& $2/3$	& 1,2	\nl
\nuc{^9Be+p} (nonresonant)\tablenotemark{*}	&$2\times2.18\ee{13}$	& $103.59$	& $2/3$	& 1	\nl
\nuc{^9Be(p,\alpha)\,^6Li} (resonant)	& $3.53\ee{16}\theta^2$		& $105.33$	& $1.367$	& 3	\nl
\nuc{^{10}B(p,\alpha)\,^7Be}	& $2.18\ee{15}$	& $120.62$	& $1/6$	& 1,2	\nl
\nuc{^{11}B(p,\alpha)2\,^4He} 	& $2.22\ee{14}$ 	& $120.95$	& $2/3$	& 1,4	\nl
\enddata

%------------------------------------------------------------------------
\tablenotetext{*}{Includes contributions from
\nuc{^9Be(p,\,^2H)2\,^4He} and nonresonant \nuc{^9Be(p,\alpha)\,^6Li}
rates.} 
\tablerefs{(1) CF88; (2) Raimann 1993; (3) Brown 1997; (4) Rauscher \&
Raimann 1996}
\end{deluxetable}

\newpage
\figcaption[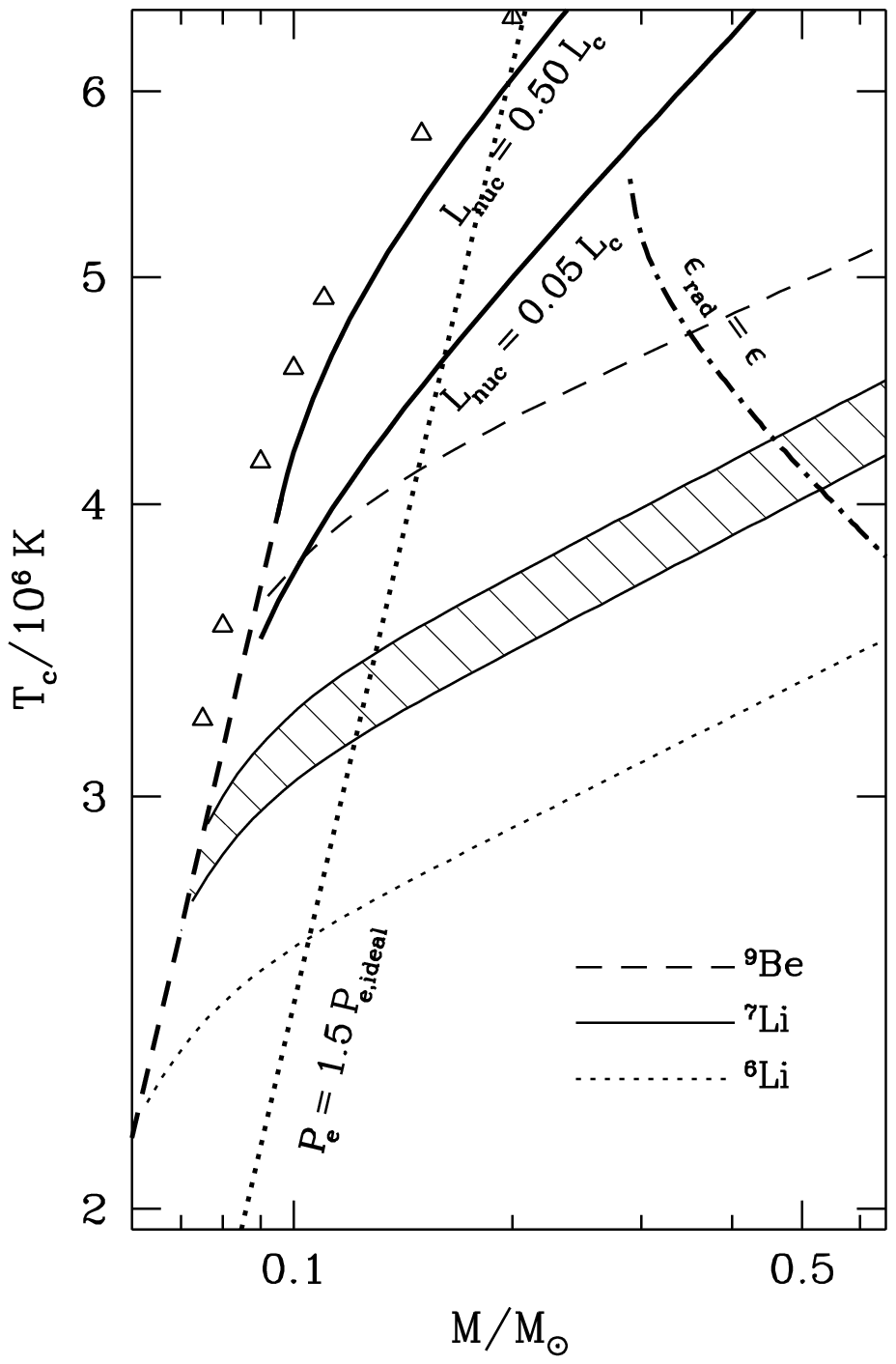]{Constraints on validity of
\nuc{^7Li}, \nuc{^6Li}, and \nuc{^9Be} calculations.  For \nuc{^7Li}
we show the temperatures at which the abundance falls from one-half to
one-hundredth its initial value ({\em shaded region\/}).  For
\nuc{^6Li} ({\em dotted line\/}) and \nuc{^9Be} ({\em dashed line\/})
we plot only the curves along which one half of the element has been
depleted.  We also show the constraints on applicability of our
depletion calculations.  Stars above the line $\epsilon_{\rm
rad}=\epsilon$ ({\em heavy dot-dashed line\/}) have radiative cores,
while the curves $L_{\rm nuc}=0.05\,L_c$ and $0.50\,L_c$ ({\em heavy
solid lines\/}) herald the arrival of the star on the main sequence.
For comparison, we show ({\em triangles\/}) central temperatures of
young ($1\Gyr$) main-sequence stars (CB97, Table 2).  Coulomb
corrections are unimportant {\em except\/} where the central
temperature is near the maximum set by degeneracy ({\em heavy dashed
line\/}), as described in \S~\protect\ref{MaximumTemperature}.  The
maximum temperature curve is essentially a curve of constant
degeneracy, as evidenced by the comparison with the curve where the
electron pressure is 1.5 times its classical value ({\em heavy dotted
line\/}).
\label{fig:MassTemperature}}

\figcaption[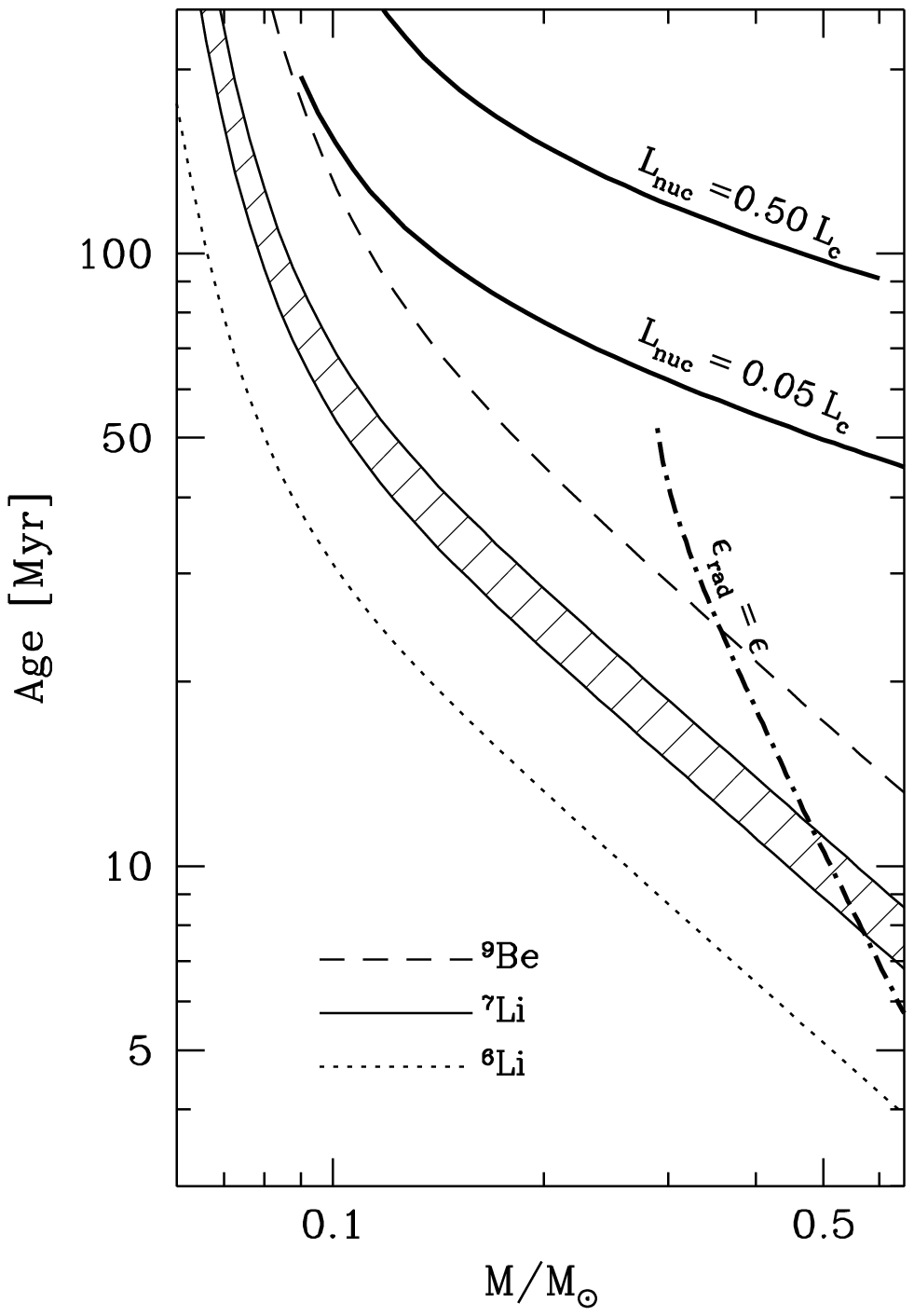]{Same as Figure
\protect\ref{fig:MassTemperature}, but for stellar age instead of
central temperature.  The main sequence, the maximum central
temperature, and the constant degeneracy lines are not shown. The
effective temperature scale $\Teff(M)$ is from the results of
\protect\cite{cha97}. 
\label{fig:MassAge}}

\figcaption[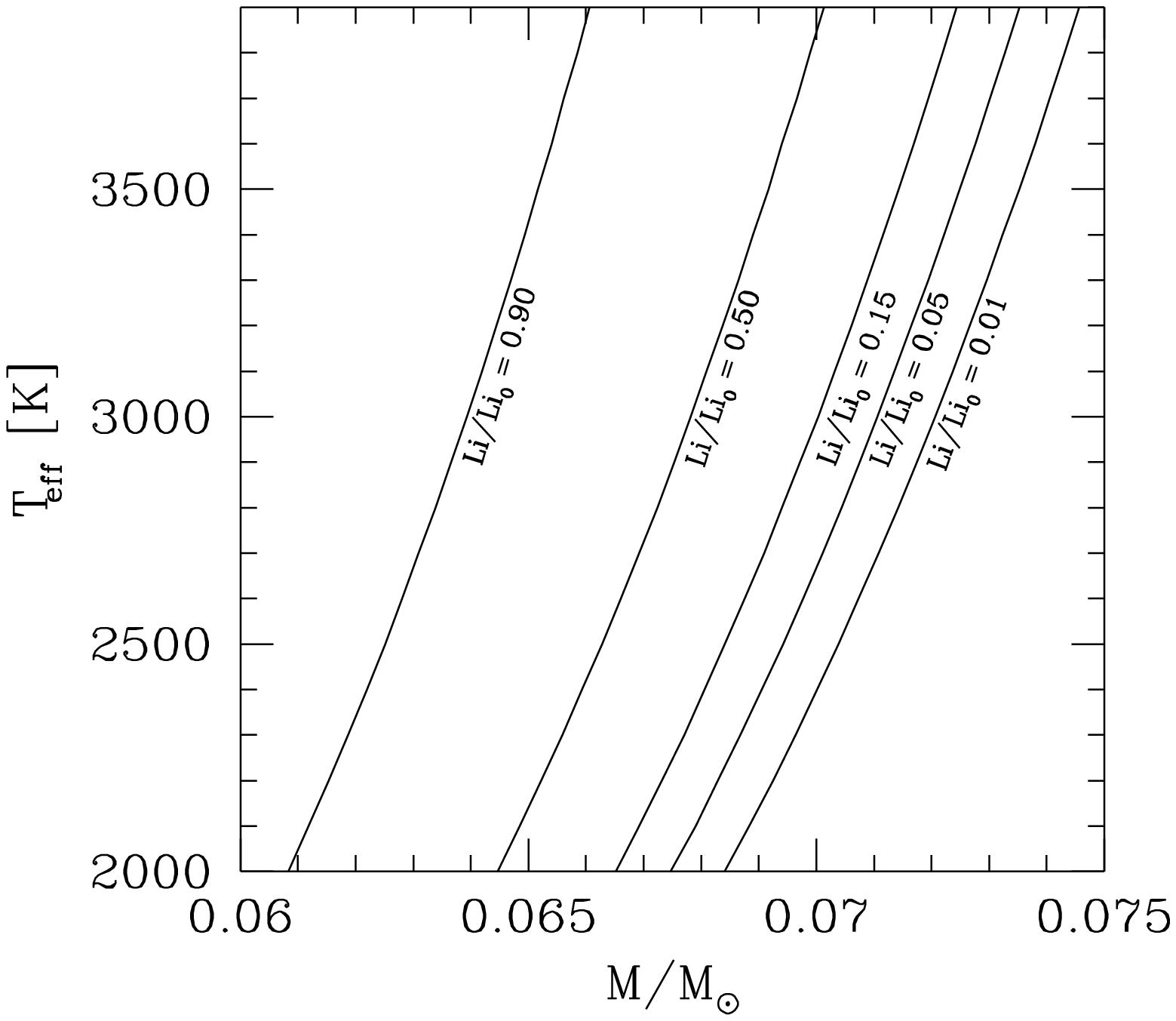]{Minimum mass, as a function of effective
temperature, for which our \nuc{^7Li} depletion calculations are
reliable.  At a given effective temperature, we show the mass for
which contraction at that $\Teff$ implies that lithium is depleted to
the given amount (\nuc{^7Li/^7Li_\circ=0.01}, 0.05, 0.15, 0.5, and
0.9) before the star reaches the maximum temperature set by degeneracy
(eq.\ [\protect\ref{eq:MaximumTemperature}]).
\label{fig:hump}}

\figcaption[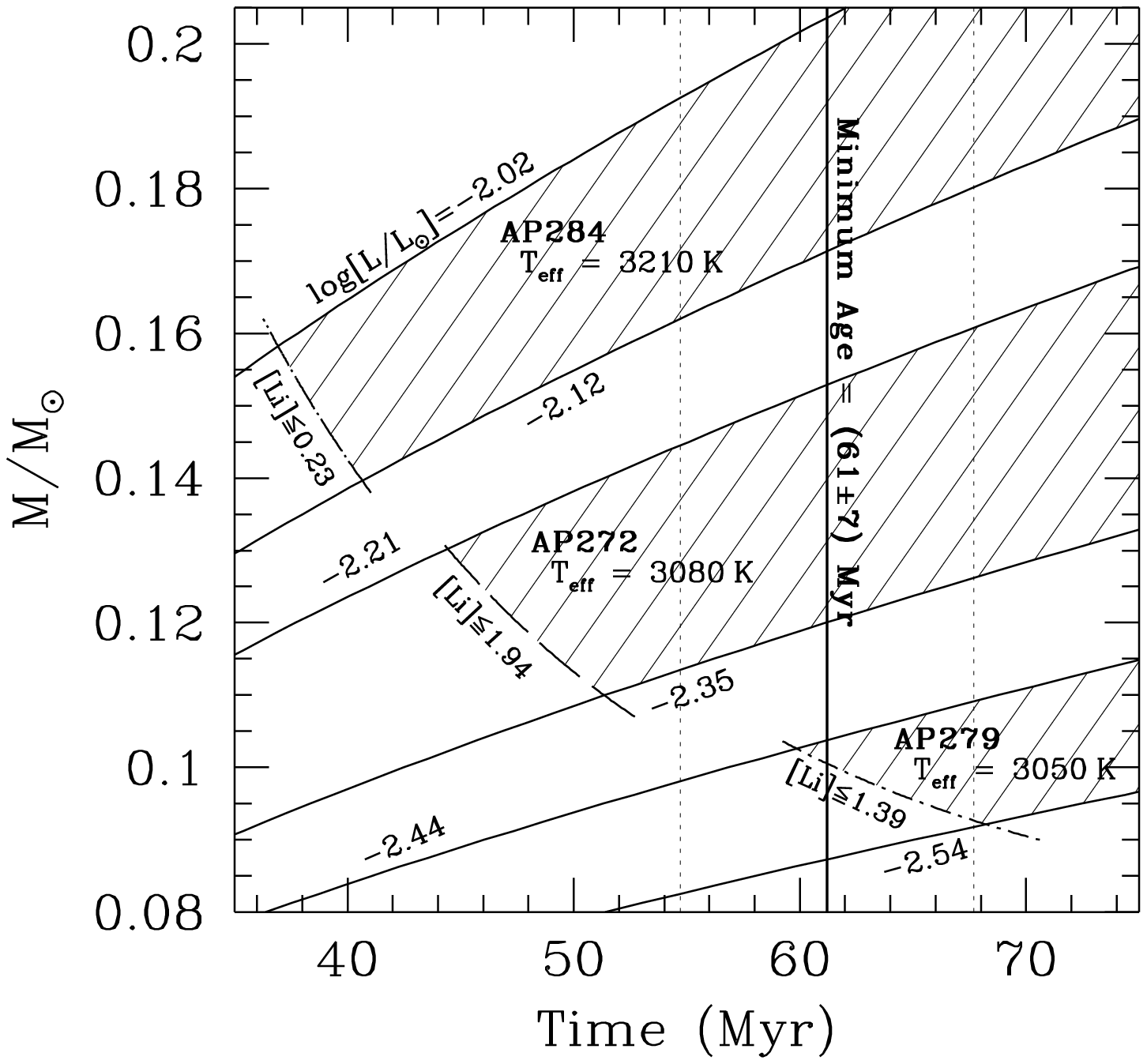]{An illustration of how the minimum age derived
from equation (\protect\ref{eq:tmin}) may be used in conjunction with
the mass-age relationship, equation (\protect\ref{eq:mass-age}), to
set a lower bound on the masses of AP284, AP272, and AP279.  We show
the minimum cluster age ({\em heavy vertical solid line}) and its
uncertainty ({\em light vertical dotted lines}).  For each star, we
plot two constant luminosity contours that correspond to the
observational uncertainty in luminosity ({\em light solid lines}) and
a contour of constant lithium abundance: AP284, $[{\rm Li}]\equiv
12+\log(N_{\rm Li}/N_{\rm H}) \le 0.23$ ({\em dot-long dashed line});
AP272, $[{\rm Li}]\le 1.94$ ({\em dashed line}); and AP279, $[{\rm
Li}]\le 1.39$ ({\em dot-short dashed line}).  These constraints fence
each star into a region of $M$--$t$ space ({\em shaded areas}).  The
intersection of the shaded region with the minimum age curve sets a
lower bound for the mass of each star.
\label{fig:aper}}

\figcaption[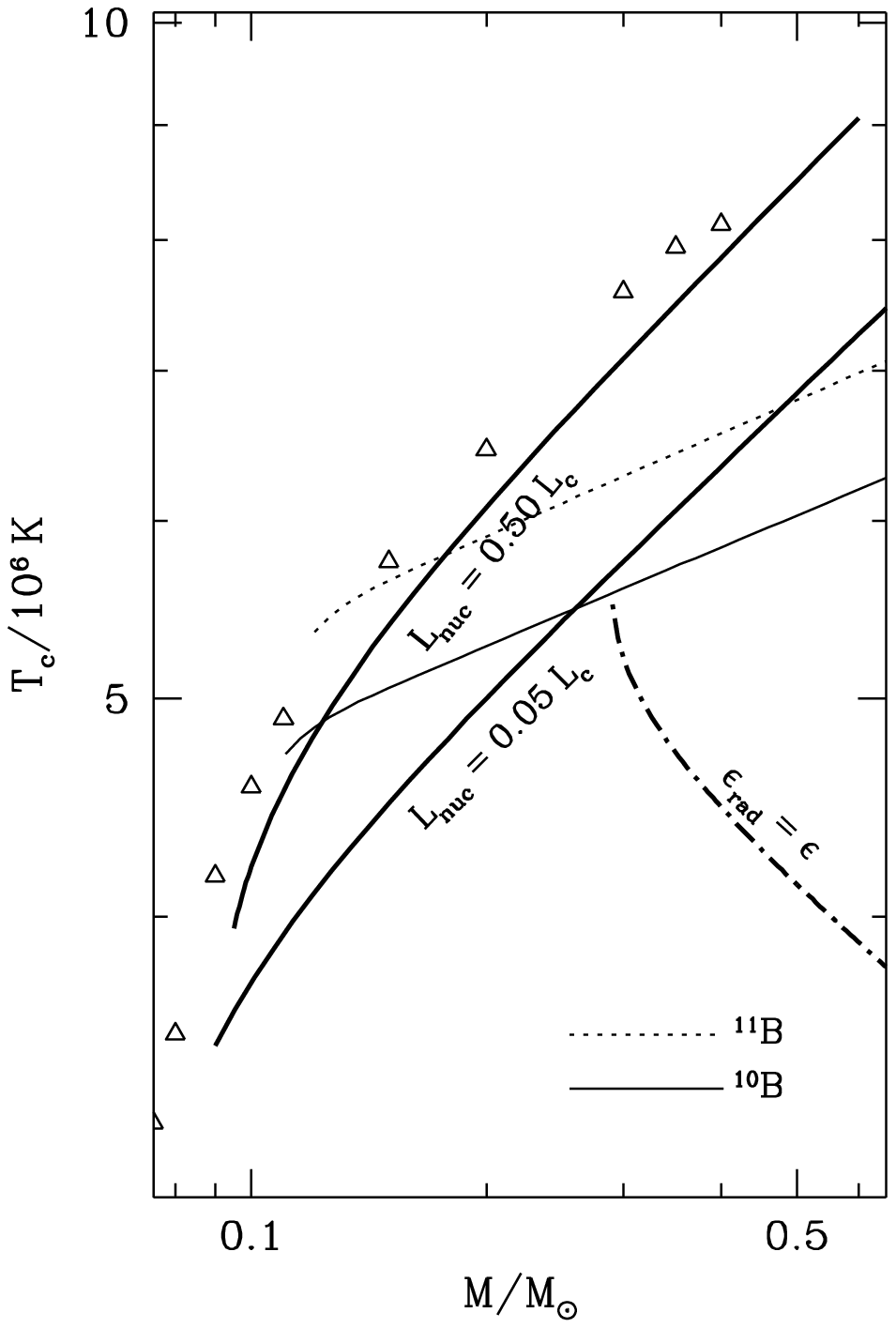]{The central temperatures at which the abundances
of \nuc{^{10}B} ({\em light solid line\/}) and \nuc{^{11}B} ({\em
light dotted line\/}) have decreased to one-half their initial values,
{\em presuming} that the star is fully convective and the p-p energy
generation is negligible. We also indicate ({\em thick solid lines\/})
where nuclear reactions contribute 5\% and 50\% of the stellar
luminosity, and also ({\em thick dot-dashed line\/}) where a radiative
region develops in the core.  For comparison, we display ({\em
triangles\/}) a $1\Gyr$ isochrone from Table 2 of Chabrier \& Baraffe
(1997).  Clearly, any treatment of boron depletion must include
hydrogen burning.  For a small range of masses about $0.1\msun$, boron
depletion occurs on the main sequence.
\label{fig:boron}}

\figcaption[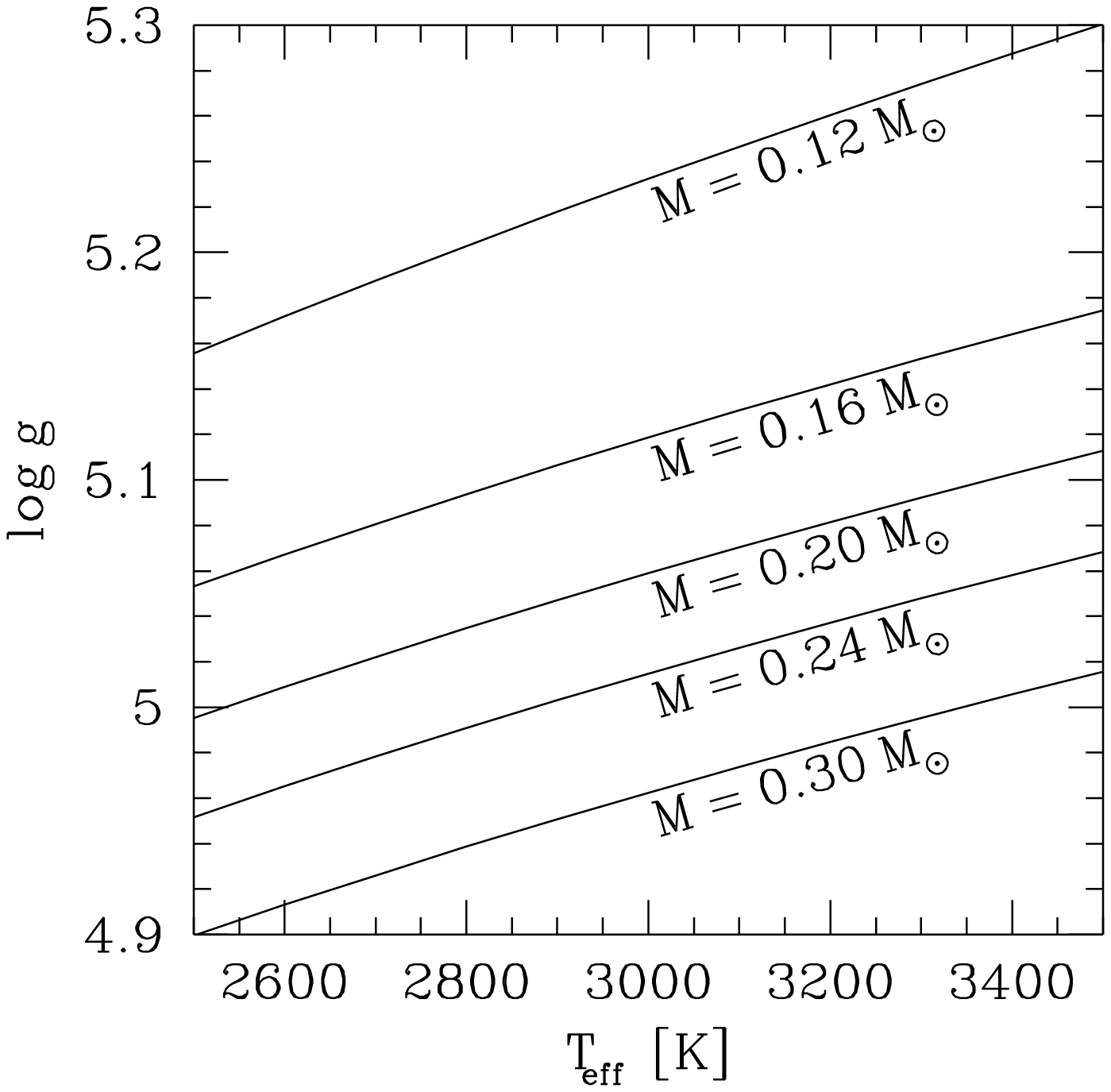]{Each line is the $\log g$--$\Teff$ relation for
a low-mass star (of mass indicated on the plot) on the Zero-Age Main
Sequence (ZAMS).  We calculate the ZAMS using the method outlined in
section \protect\ref{coreconv}.
\label{fig:gTeff}}

\end{document}